\documentclass[journal]{IEEEtran}
\ifCLASSOPTIONcompsoc
  \usepackage[nocompress]{cite}
\else
  \usepackage{cite}
\fi
\ifCLASSINFOpdf
\else
\fi
\usepackage{cite}
\usepackage{amsmath,amssymb,amsfonts}
\usepackage{graphicx}
\usepackage{textcomp}
\usepackage{xcolor}
\usepackage{comment}
\usepackage{enumitem}
\usepackage{multirow}
\usepackage{booktabs}
\usepackage{soul}
\usepackage{amsmath,bm}
\usepackage{xspace}
\usepackage{pifont}
\usepackage{lipsum}
\usepackage{graphics}
\usepackage{algorithm}
\usepackage{algpseudocode}
\usepackage{amsmath}
\usepackage{graphicx}
\usepackage{subfigure}
\usepackage{pifont}
\usepackage{mathtools}
\usepackage{hyperref}
\usepackage{ragged2e}

\usepackage{cite}
\usepackage{amsmath,amssymb,amsfonts}
\usepackage{graphicx}
\usepackage{textcomp}
\usepackage{xcolor}
\usepackage{CJK}
\usepackage{indentfirst}
\usepackage{flushend}
\renewcommand{\raggedright}{\leftskip=0pt \rightskip=0pt plus 0cm}

\renewcommand{\raggedright}{\leftskip=0pt \rightskip=0pt plus 0cm} 
\def\BibTeX{{\rm B\kern-.05em{\sc i\kern-.025em b}\kern-.08em
		T\kern-.1667em\lower.7ex\hbox{E}\kern-.125emX}}

\begin{document}
	
\title{Energy-Efficient Accelerator Design for Deformable Convolution Networks}

\author{
Dawen~Xu, 
Cheng~Chu, 
Cheng~Liu,
Ying~Wang, \\
Huawei Li, ~\IEEEmembership{Senior~Member,~IEEE, } 
Xiaowei Li,
Kwang-Ting Cheng, ~\IEEEmembership{Fellow,~IEEE} 

\IEEEcompsocitemizethanks{
{\IEEEcompsocthanksitem Corresponding author: Cheng Liu\\}
{\IEEEcompsocthanksitem Dawen Xu and Cheng Chu are with both Hefei University of Technology, Anhui, 230009 and Institute of Computing Technology (ICT), Chinese Academy of Sciences (CAS), Beijing, China, 100180.\\}
{\IEEEcompsocthanksitem Cheng Liu, Ying Wang, Huawei Li and Xiaowei Li are with ICT, CAS, Beijing, China, 100180. (E-mail: liucheng@ict.ac.cn)\\}
{\IEEEcompsocthanksitem Kwang-Ting Cheng is with Department of Computer Science and Engineering, The Hong Kong University of Science and Technology, Hong Kong.(E-mail:timcheng@ust.hk)\protect\\}
}


}

\IEEEtitleabstractindextext{%
	
\begin{abstract}
\raggedright Deformable convolution networks (DCNs) proposed to address the image recognition with geometric or photometric variations typically involve deformable convolution that convolves on arbitrary locations of input features. The locations change with different inputs and induce considerable dynamic and irregular memory accesses which cannot be handled by classic neural network accelerators (NNAs). Moreover, bilinear interpolation (BLI) operation that is required to obtain deformed features in DCNs also cannot be deployed on existing NNAs directly. Although a general purposed processor (GPP) seated along with classic NNAs can process the deformable convolution, the processing on GPP can be extremely slow due to the lack of parallel computing capability. 
  
To address the problem, we develop a DCN accelerator on existing NNAs to support both the standard convolution and deformable convolution. Specifically, for the dynamic and irregular accesses in DCNs, we have both the input and output features divided into tiles and build a tile dependency table (TDT) to track the irregular tile dependency at runtime. With the TDT, we further develop an on-chip tile scheduler to handle the dynamic and irregular accesses efficiently. In addition, we propose a novel mapping strategy to enable parallel BLI processing on NNAs and apply layer fusion techniques for more energy-efficient DCN processing. According to our experiments, the proposed accelerator achieves orders of magnitude higher performance and energy efficiency compared to the typical computing architectures including ARM, ARM+TPU, and GPU with 6.6\% chip area penalty to a classic NNA.
\end{abstract}

}

\maketitle
\IEEEdisplaynontitleabstractindextext
\IEEEpeerreviewmaketitle

\section{Introduction} \label{sec:intro}
Deformable convolution network (DCN) \cite{dai2017deformable}, a new category of neural networks, is proposed to address the neural network model accuracy degradation caused by geometric and photometric variations such as lighting and rotation occurred in many practical applications like medical imaging. DCNs typically sample arbitrary locations of the input features for the convolution such that the objects with different scales or deformation can be captured. The sampling patterns of deformable convolution can be learned and calculated using an additional convolution layer. With the unique deformable convolution, DCNs have shown superior performance on many vision tasks such as object detection \cite{dai2017deformable} \cite{cao2019object} \cite{8953851} \cite{8901912}, semantic segmentation \cite{dai2017deformable} \cite{deng2019restricted} \cite{chen2018encoder} \cite{8953750} and classification \cite{diao2020multi} \cite{ludeformsketchnet}\cite{lai20213d}. For instance, the authors in \cite{dai2017deformable} demonstrated that the prediction accuracy of the proposed DCN increases from 70\% to 75\% on the image semantic segmentation dataset (CityScapes). Significant prediction accuracy improvement is also observed in human motion recognition task \cite{sun2018integral} \cite{weng2018deformable}, action detection task \cite{bertasius2018object} \cite{mac2018locally} and intelligent medical monitoring and treatment \cite{9206944} \cite{8999148}.  

\begin{figure*}
	\center{\includegraphics[width=0.85\linewidth]{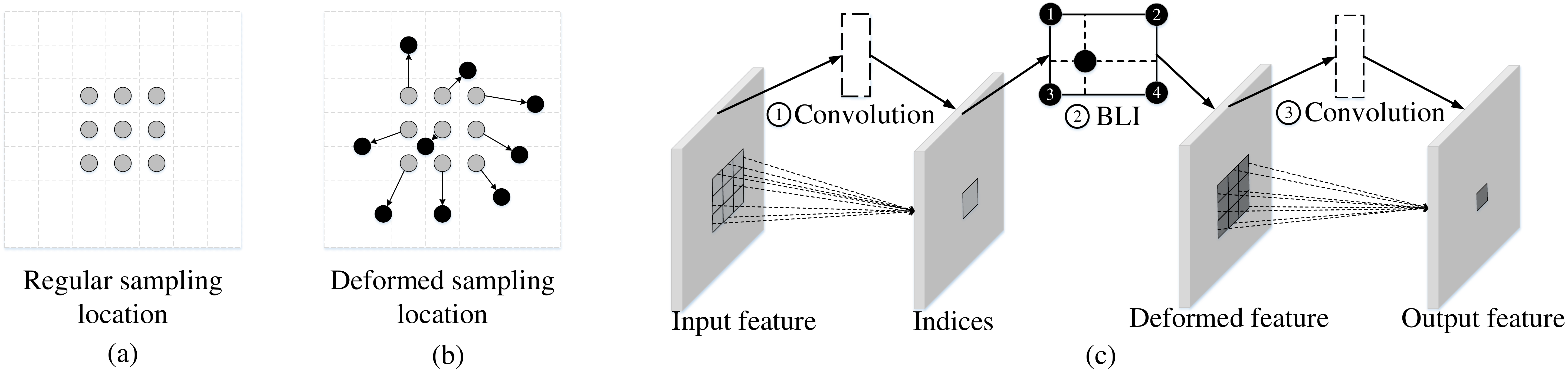}}
    \caption{Deformable convolution (a) regular sliding window in a standard convolution, (b) irregular sampling of a deformable convolution, (c) deformable convolution processing.}
\label{fig:dcnn}
\vspace{-0.5em}
\end{figure*}

Despite the great advantages, each deformable convolution operation in DCNs needs additional convolution-based index\footnote{'offset' is used to represent the relative distance to sliding window positions of a standard convolution in many DCNs. However, it is inconvenient to retrieve the features with the random offsets in hardware and offsets need to be converted to indices of the features instead. While the conversion is trivial, we use index in both the algorithm description and hardware description to make the notation consistent across the paper.} calculation and bilinear interpolation (BLI) to obtain deformed features other than a standard convolution, so it is both computing- and memory-intensive and requires intensive acceleration for widespread deployment. Nevertheless, DCNs can not be deployed on conventional neural network accelerators mainly from the following two aspects. First, it convolves on arbitrary locations of the input features instead of fixed sliding windows as depicted in \autoref{fig:dcnn}. The locations i.e. the indices to the input features are generated at runtime and they will cause both dynamic and irregular accesses to the memory, which can not be fitted to conventional neural network accelerators targeting at regular memory accesses and data flows. Second, DCNs have a standard convolution to calculate the indices, but the calculated indices are usually not integers and can not be used to retrieve the feature data directly. Typically, BLI algorithm is utilized to approximate the features with the nearest original input features. This step is also not supported in conventional neural network accelerators. An intuitive solution to execute DCNs is to conduct the deformable convolution on a general purposed processor (GPP) which is usually seated with a neural network accelerator while deploying the rest of the normal neural network operations in DCNs on a neural network accelerator. However, GPPs especially the embedded processors with limited parallel processing capability is inefficient for the bilinear interpolation, and a large number of irregular memory accesses and data transfers between the GPPs and the neural network accelerator also inhibit the DCN execution efficiency dramatically. 

Recently, there are also works proposed to revise the DCN models to fit existing neural network accelerators. The authors in \cite{Algorithm} proposed to replace the bilinear interpolation algorithm with a simple rounding strategy and restrict the sampling locations to avoid dynamic memory access induced buffering problems. Similarly, the authors in \cite{efficient} also proposed to modify the DCN models to reduce the receptive field size substantially so that the sampling locations are limited to a small region, which avoids the dynamic memory accesses across the whole input feature map. Although these approaches are demonstrated to be effective on existing neural network accelerators with minor model accuracy, it essentially poses hardware constraints to the model design and particularly limits its use on scenarios that are sensitive to the model accuracy loss. In addition, it requires time-consuming retraining and training data that may not be applicable to the end users. Thereby, we investigate the computing of the DCNs and seek to implement the entire unmodified deformable convolution on top of a unified neural network accelerator directly.

To implement the entire DCNs on a unified accelerator and reuse the conventional neural network accelerator as much as possible, we revisit a typical neural network accelerator architecture mainly for the new irregular feature sampling and the BLI required by DCNs. For the dynamic and irregular feature accesses, we observe that the input data required by the deformable convolutional output is imbalanced and some of the input features are utilized more than the others. More details can be found in Section \ref{sec:observation}. With this observation, we propose to divide the input and output features into smaller tiles and build a tile dependency table (TDT) that keeps a record of all the required input tile IDs of each output tile with runtime tracking. On top of the TDT, we further schedule the output tile execution such that the buffered tiles are reused as much as possible and the overall memory access efficiency can be improved. For the BLI, we convert it to multiple small vector-based dot production which can be mapped in parallel to the 2D computing array in typical neural network accelerators efficiently with a weight stationary data flow \cite{chen2016eyeriss}. In addition, we fuse the BLI and the following convolution to further reduce the intermediate data transmission between on-chip buffers and the external memory. With the proposed redesigning on top of a conventional neural network accelerator, the entire deformable convolution can be implemented on the revisited accelerator efficiently. According to our experiments on a set of DCNs, it achieves orders of magnitude higher performance and energy efficiency when compared to typical computing architectures including ARM, ARM+TPU, and GPU while it incurs only minor hardware resource consumption compared to a conventional neural network accelerator. 

The rest of the paper is organized as follows. In Section \ref{sec:background}, we introduce the typical deformable convolutional networks and formulate them into a unified computing model. Meanwhile, we brief prior works about neural network accelerator redesigning for new types of neural network models. In Section \ref{sec:observation}, we characterize the computing patterns and memory accesses of deformable convolution. In Section \ref{sec:overview}, we show the detailed design and optimizations of an accelerator for DCNs on top of a classical neural network accelerator. In Section \ref{sec:result}, we evaluate the performance and energy efficiency of the accelerator and compare it with typical embedded computing platforms. In Section \ref{sec:conclusion}, we conclude this work.

\section{Background and Related Work} \label{sec:background}
\subsection{Deformable convolutional networks}
Deformable convolution may sample arbitrary locations of the input features for convolution such that it can capture the objects with scale or transformation. The unique feature makes it attractive in visual recognition tasks with geometric variations such as lighting and rotation. There have been many deformable convolution architectures proposed recently \cite{dai2017deformable, mac2018locally}. They typically include two standard convolution operations. The first convolution calculates the indices for the input feature sampling while the second convolution convolves on the sampled features. Usually, a BLI operation is used to bridge the two convolution operations. It approximates the input features based on the non-integer indices generated by the first convolution and provides the features as inputs to the second convolution. The structures of DCNs mainly differ on the index reuse and can be roughly divided into two categories. The first category of DCNs has a unique index for each data in the feature plane and the indices are reused across the different channels \cite{mac2018locally}. The second category of DCNs also has the indices shared across the feature channels, but it has a unique index for each data in each convolution window \cite{dai2017deformable}. Basically, the same data in the feature plane has different indices when it is located in different convolution windows. The second category of DCNs requires a larger convolution to calculate more sampling indices and produces more deformed features than the first category of DCNs. The two categories of DCNs are abbreviated as DCN-I and DCN-II respectively.

\subsection{Unified deformable convolution model} \label{sec:dcn-model}
\begin{figure}
	\center{\includegraphics[width=0.8\linewidth]{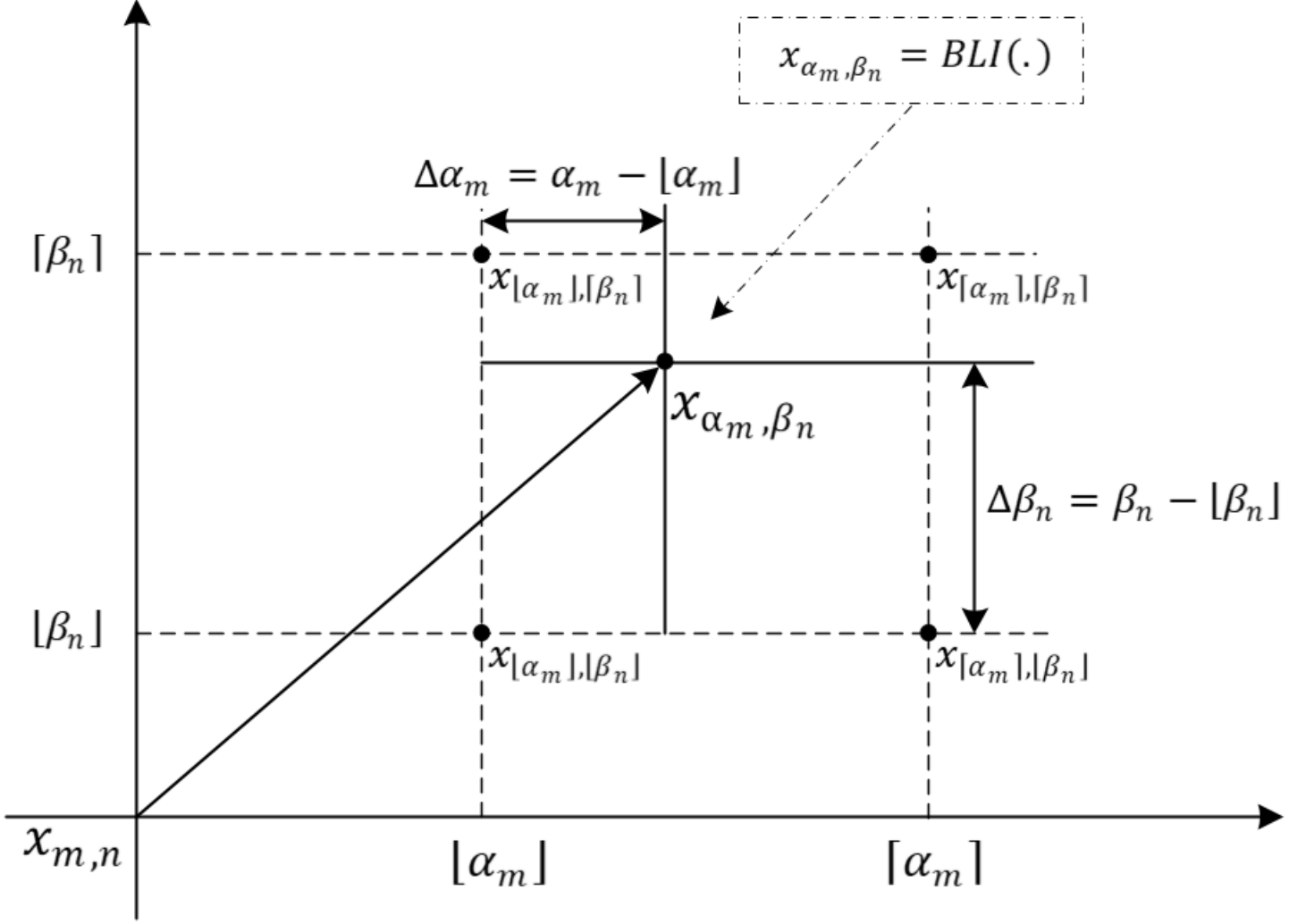}}
    \caption{Bilinear interpolation}
\label{fig:BLI}
\vspace{-0.5em}
\end{figure}

The different deformable convolution can be represented in a unified model as formulated in Equation \ref{eq:first-conv}-\ref{eq:second-conv}. Basically, it has a convolution to determine the deformed locations or indices of the input features in the first step. This step is formulated in Equation \ref{eq:first-conv} where $c$, $l$, $i$, $j$ refer to input channel index, output channel index, kernel offset in $X$ dimension and kernel offset in $Y$ dimension respectively. Since the indices can be different when the feature data is in different positions of the convolution kernel, we have \bm{${y_L}$}, a vector of planar coordinates with length $L=2 \times K \times K$ where $K$ is the kernel size to represent the indices. Suppose $\alpha_{m}$ and $\beta_{n}$ are the corresponding coordinates in $X$ axis and $Y$ axis respectively. $\alpha_{m}$ and $\beta_{n}$ are not integers, so they can not be used to retrieve the input features directly for the deformable convolution. To that end, a bilinear interpolation approach is utilized to calculate the deformed features using the neighboring features around the location ($\alpha_{m}$, $\beta_{n}$) in the second step. The calculated feature $x_{c,\alpha _{m},\beta _{n}}^{'}$ can be obtained using Equation \ref{eq:bli} where $BLI(.)$ refers to the bilinear interpolation function, $\Delta \alpha _{m}=\alpha _{m}-\left \lfloor \alpha _{m} \right \rfloor$ and $\Delta \beta _{m}=\beta _{m}-\left \lfloor \beta _{m} \right \rfloor$. A vivid description of the BLI function can be found in \autoref{fig:BLI}. When the input features are retrieved and organized according to the deformed indices, the deformable convolution can be obtained using a standard convolution over the reorganized features as shown in Equation \ref{eq:second-conv} in the third step. 

\begin{equation}
\label{eq:first-conv}
\begin{aligned}
    y_{L}=\sum_{c}^{}\sum_{i,j}^{}w_{c,l,i,j}\cdot x_{c,i,j}+b_{l}
    \end{aligned}
\end{equation}

\begin{equation}
\label{eq:bli}
\begin{aligned}
    x_{c,\alpha _{m},\beta _{n}}^{'}&= F_{BLI}( x_{\left \lfloor \alpha _{m}\right \rfloor,\left \lfloor \beta _{n} \right \rfloor},x_{\left \lfloor \alpha _{m} \right \rfloor,\left \lceil \beta _{n} \right \rceil},x_{\left \lceil \alpha _{m} \right \rceil,\left \lfloor \beta _{n} \right \rfloor},\\
    &\;\;\;\;x_{\left \lceil \alpha _{m} \right \rceil,\left \lceil \beta _{n} \right \rceil},\Delta \alpha _{m},\Delta \beta _{n} )\\
    &=\left ( 1-\Delta \alpha _{m} \right )\left ( 1-\Delta \beta _{n} \right )x_{\left \lfloor \alpha _{m}\right \rfloor,\left \lfloor \beta _{n} \right \rfloor}\\
    &\;\;\;\;+\left ( 1-\Delta \alpha _{m} \right )\Delta \beta _{n}x_{\left \lfloor \alpha _{m} \right \rfloor,\left \lceil \beta _{n} \right \rceil}\\
    &\;\;\;\;+\Delta \alpha _{m}\left ( 1-\Delta \beta _{n} \right )x_{\left \lceil \alpha _{m} \right \rceil,\left \lfloor \beta _{n} \right \rfloor}\\
    &\;\;\;\;+\Delta \alpha _{m}\Delta \beta _{n}x_{\left \lceil \alpha _{m} \right \rceil,\left \lceil \beta _{n} \right \rceil}
\end{aligned}
\end{equation}

\begin{equation}
\label{eq:second-conv}
    y_{L^{'}}^{'}=\sum_{c}^{}\sum_{m,n}^{}w_{c,l^{'},m,n}^{'}\cdot x_{c,m,n}^{'}+b_{l^{'}}
\end{equation}

\subsection{Neural Network Accelerator Redesigning}
The great success of neural networks in massive domains of applications inspires considerable efforts devoted to developing neural network accelerators \cite{albericio2016cnvlutin} \cite{bromberger2016fpga} \cite{wang2016deepburning} \cite{zhang2018caffeine} \cite{xu2021hyca} \cite{xu2020hybrid}. In spite of the notable efforts, newer network operations proposed for higher performance may go beyond the capability of the existing neural network accelerators. Although it is usually possible to offload the unsupported operations to the attached GPPs while leaving the rest of the conventional neural network operations on the accelerator, the performance of the co-designed implementation may drop dramatically due to the massive data communication between GPP and the accelerator. Also complex operations offloaded to GPPs may still become the performance bottleneck due to the insufficient computing capability of GPP and degrade the overall neural network execution. Thereby, the neural network accelerators are usually redesigned to meet the requirements of the new neural network operations on top of the existing neural network accelerators. For instance, unified neural network accelerators are also proposed to perform deconvolution used in generative adversarial neural networks other than the conventional convolution\cite{xu2018fcn}\cite{yazdanbakhsh2018ganax}\cite{yan2018gna}\cite{zhang2017design}. A novel accelerator is developed to support 3D neural networks in \cite{hegde2018morph}\cite{wang2017enhanced}\cite{fan2018reconfigurable}. The authors in \cite{RECOIN} proposed to add a bilinear interpolation calculation module to an existing ReRAM neural network accelerator to enable in-situ DCN calculation. Inspired by prior works, we also try to convert the deformable convolution to operations that can be mapped to the neural network accelerator efficiently with minor hardware modification such that the whole deformable convolution can be executed on the neural network accelerator efficiently. 
\section{Observation} \label{sec:observation}
Since deformable convolution has many irregular memory accesses involved which dramatically affect the processing efficiency, we investigate the memory accesses of a typical deformable convolution in this section. We take the third convolution layer of VGG16 as the basis of a typical deformable convolution operation. As the memory accesses vary on different inputs, we randomly selected 2000 images from ImageNet and averaged the memory accesses for the investigation.

\begin{figure}
\setlength{\abovecaptionskip}{0pt}
\setlength{\belowcaptionskip}{-12pt}
	\center{\includegraphics[width=0.95\linewidth]{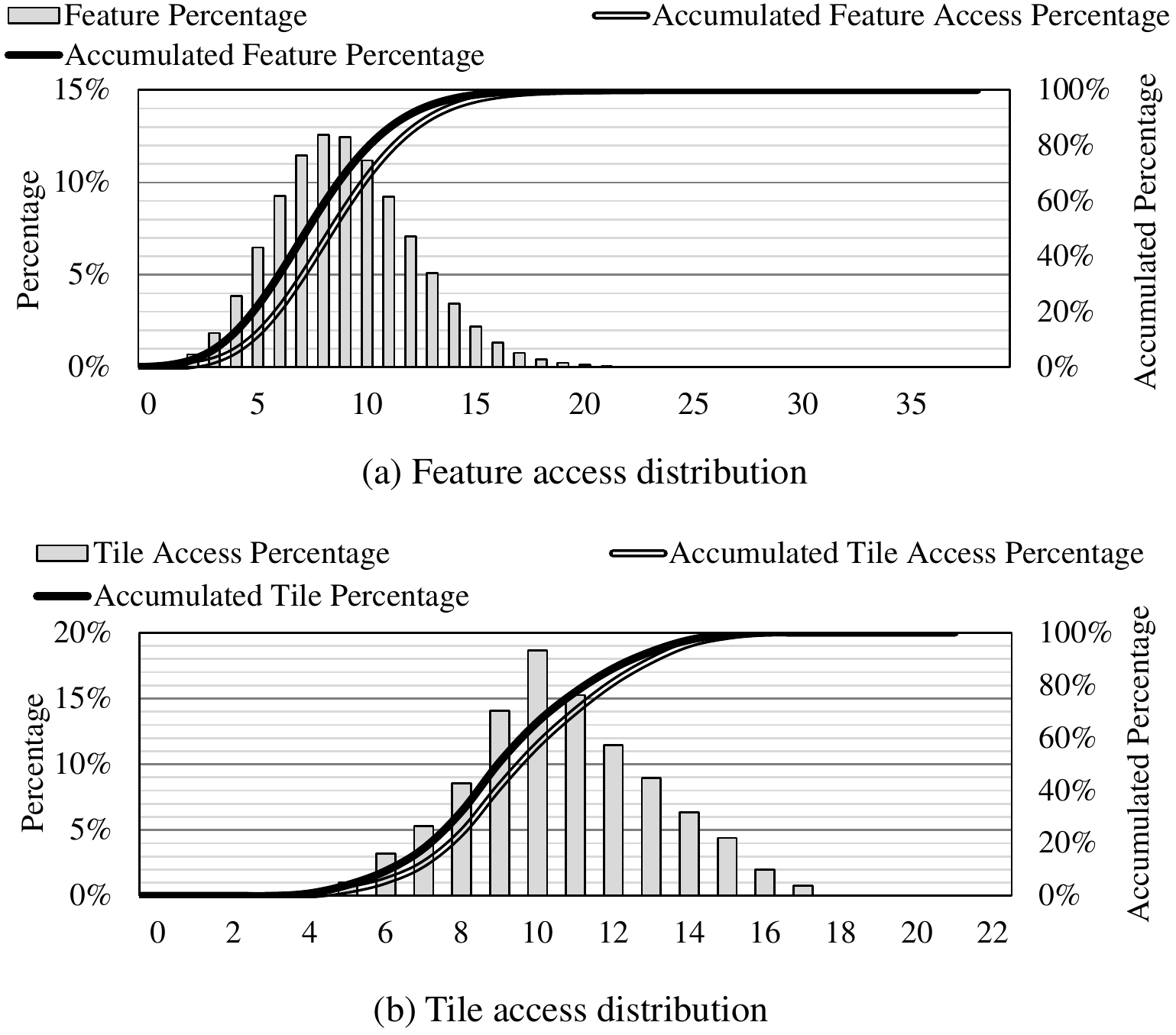}}
    \caption{Memory access characterization of the deformable convolution}
\label{fig:data}
\vspace{-0.5em}
\end{figure}

Figure \ref{fig:data} (a) shows the distribution of the input feature accesses. Unlike standard convolution that usually has nearly uniform utilization of the input features, deformable convolution has distinct utilization of the different input features. With the $3 \times 3$ kernel used in the convolution, each input feature will be utilized around 9 times. For the corresponding deformable convolution, the access distribution shows dramatic difference. Around 15\% of the features will be utilized by more than 12 times, which take up around 25\% of the feature accesses. In contrast, more than 22\% of the features are utilized less than 6 times. While the neural networks can usually be tiled and the memory accesses can be performed in the granularity of a tile, we further analyze the input feature tile access distribution. In this experiment, we had both the input feature map and the output feature map divided into 25 tiles as an example. The tile access distribution is shown in Figure \ref{fig:data} (b). It can be observed that the input tile reuse still shows 
notable variation. 

In summary, the input features are not evenly accessed, so some of the input features are more likely to be reused than the other. Given limited on-chip buffers, scheduling the order of the output feature calculation and optimizing the order of the input feature accesses can potentially improve the on-chip buffer utilization and remains demanded for higher performance and energy efficiency of the DCN processing.

\section{DCN accelerator Architecture} \label{sec:overview}
\subsection{Overall Accelerator Architecture}
Deformable convolution is the major barrier that hinders the deployment of DCNs on existing neural network accelerators. Thereby, taming the deformable convolution to the existing neural network accelerators is key to accelerate DCNs. As formulated in Section \ref{sec:dcn-model}, deformable convolution consists of three processing stages including convolution, BLI, and convolution. Since the convolution operations in DCNs can be deployed on existing neural network accelerators directly, the major challenge of DCN acceleration is to optimize the BLI operation that samples the input features according to the irregular indices calculated with the first convolution in DCNs and conducts the BLI calculation based on the sampled input features. Since the indices depend on the input features and thus change at runtime, the BLI sampling leads to complicate memory accesses. To address this dynamic and irregular memory access problem, we have BLI divided into tiles and track the tile dependency at runtime with a tile dependency table (TDT). On top of the TDT, we have a tile scheduler to decide the order of the output tile execution and the input tile loading at the same time such that the tiles loaded to the on-chip buffers can be fully utilized and the memory accesses to the external memory can be reduced. As for the BLI calculation, we reorganize it to multiple small vector-based dot production and have the processing performed in parallel on top of the 2D computing array in the neural network accelerator for the sake of higher performance.

\begin{figure}[tb]
	\center{\includegraphics[width=0.95\linewidth]{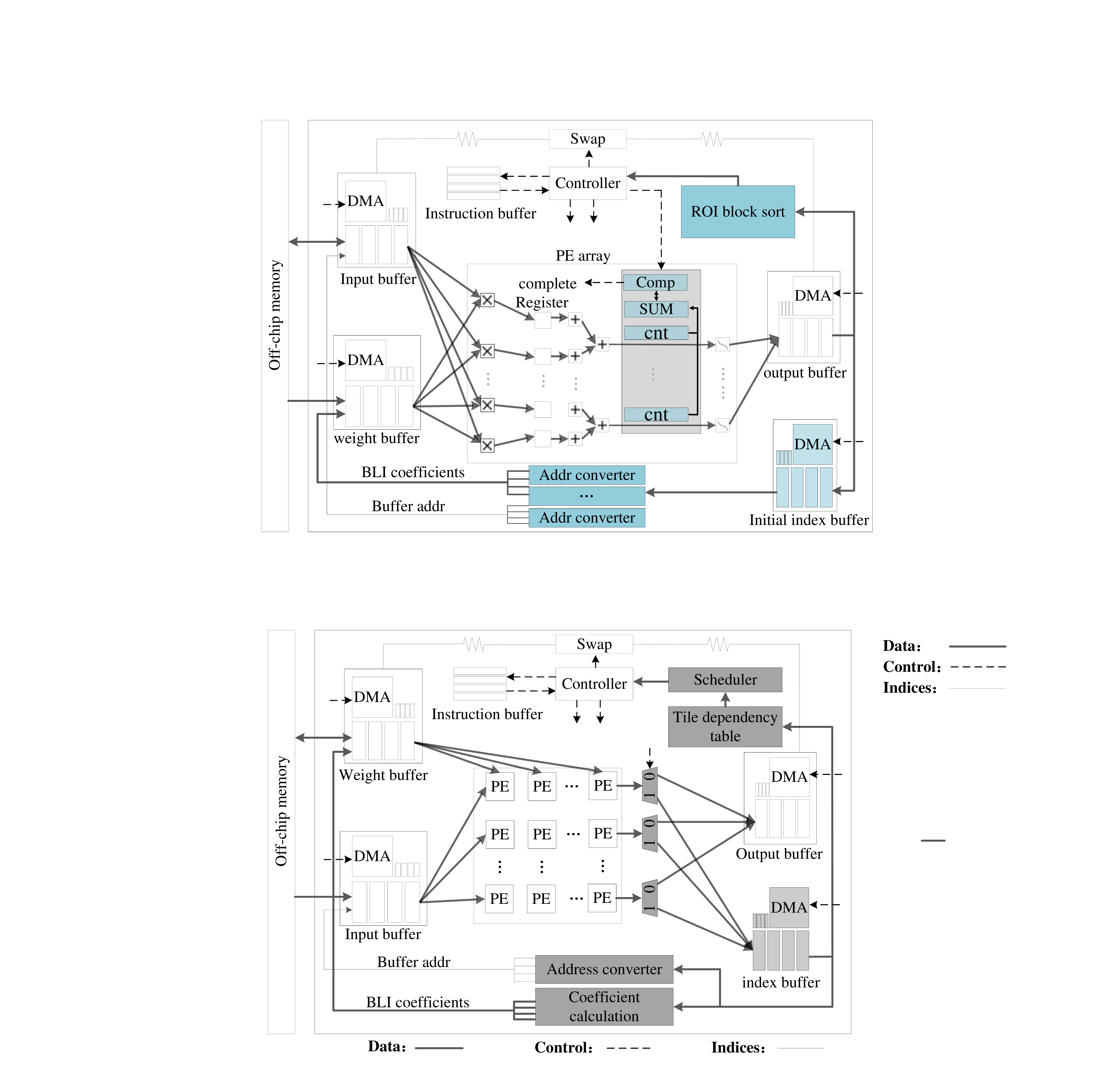}}
    \caption{DCN accelerator overview. The blocks filled with grey are added specifically for the deformable convolution processing while the rest of the blocks remain the same with a conventional neural network accelerator with 2D computing array.}
\label{fig:arch}
\vspace{-0.5em}
\end{figure}

The proposed DCN accelerator architecture is shown in \autoref{fig:arch}. The components without filling any color belong to a classic neural network accelerator while the rest of the components filled with grey are designed specially for the deformable convolution. The entire accelerator is generally controlled with a sequence of instructions compiled from the target neural networks. The instructions are stored in the instruction buffer and decoded at runtime to generate control signals for the entire accelerator. With the controlling, neural network operations are mapped and executed on the regular 2D computing array. For the convolution operations, weights from different filters are streamed to the different columns of the computing array from top to bottom in parallel while input features are streamed from left to right in different rows according to the output stationary data flow proposed in \cite{chen2016eyeriss}. 

For the deformable convolution, it is divided into three dependent operations. The first convolution operation starts when inputs and weights are ready in the on-chip buffers. The outputs of the first convolution are indices and will be utilized to sample from the input features. Since they are not integers and can not be utilized to retrieve the input features directly, we have them stored in an index buffer and have an address converter to obtain the four neighboring integer indices. Meanwhile, BLI coefficients as proposed in Section \ref{sec:dcn-model} are generated with a coefficient calculation block at the same time. To this end, we can start the BLI calculation with the retrieved input features and BLI coefficients by taking advantage of the 2D computing array of the accelerator. The output of the BLI are essentially deformed features and will be utilized as inputs of the second convolution. While the address conversion and the BLI calculation are conducted in pipeline manner, the output buffer and the index buffer are separated in case of conflicts. After the BLI calculation, the deformed features will be stored and swapped as inputs of the computing array for the second convolution. When the features or weights exceed the on-chip buffers, BLI and the second convolution need to be tiled and fused to avoid intermediate data exchange through the external memory. In addition, we have a TDT to keep a record of the tile dependency based on the generated indices and have a runtime tile scheduler to optimize the output tile execution and input tile loading ordering based on the TDT to enhance the on-chip data reuse and memory access efficiency.

\subsection{BLI Implementation}\label{subsection:block_sch}
To implement BLI, we have an address converter module to convert the original non-integer indices to neighboring integer addresses of the input feature buffer and a coefficient calculation block to produce the BLI coefficients at the same time. Each feature requires four coefficients $\eta, \mu, \theta, \gamma$ and they can be calculated with Equation \ref{eq:bli}. The BLI coefficients are stored in the weight buffer while the converted buffer addresses are used to retrieve features from the the input buffer directly. The retrieved features and coefficients read from the weight buffer will then be fed to the 2D computing array following a standard weight stationary data flow \cite{chen2016eyeriss} for the BLI computing. Details of the processing will be illustrated in the rest of this section.

\begin{figure}[tb]
	\center{\includegraphics[width=0.80\linewidth]{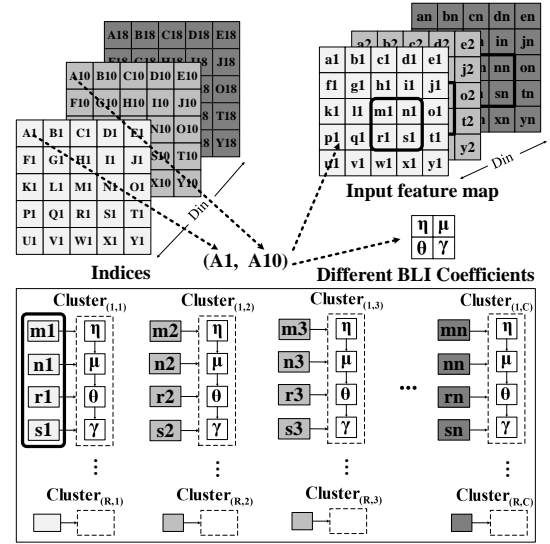}}
    \caption{BLI mapping strategy}
\label{BLI-mapping}
\vspace{-0.5em}
\end{figure}

\begin{figure}[tb]
	\center{\includegraphics[width=0.95\linewidth]{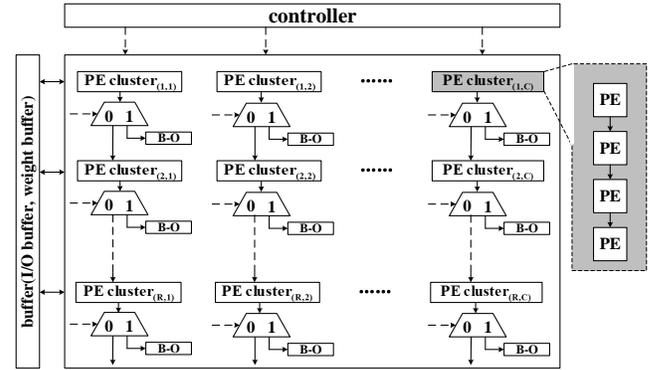}}
    \caption{Clustered PE array for parallel BLI calculation. The design is built on top of a conventional 2-D computing array in typical neural network accelerators. When the MUXs select 0, it is configured to be a normal 2-D computing array and can be used for standard convolution. When the MUXs select 1, each PE cluster in the design can be used for an BLI output calculation. }
\label{pe-modification}
\vspace{-0.5em}
\end{figure}

\textbf{BLI Mapping} Each deformed feature depends on four neighboring input features and it can be viewed as a vector-based dot production. One of the vector includes four BLI coefficients and the other vector includes four neighboring input features as shown in Figure \ref{BLI-mapping}. Each deformed feature processing can be mapped to four neighboring PEs organized as a cluster with a weight-stationary data flow. Since weights of the BLI are shared among the different input feature channels, they can be distributed to the PEs in the same cluster and broadcast to different clusters. The corresponding four input features in the same channel will be streamed in parallel for the multiplication among PEs in the same cluster. Features from different channels will be distributed to the different clusters of the computing array for higher throughput, but additional wires from wide input feature buffers to the PEs across the computing array are required accordingly. The clustered computing array on top of the original 2D regular computing array is shown in Fig.\ref{pe-modification}. Unlike the output of the conventional computing array that are aligned in column, the clustered computing array output are extracted and aligned in cluster. Thereby, a MUX is added to the output port of each PE cluster to extract the output from the computing array when BLI is mapped.

\begin{figure}[tb]
	\center{\includegraphics[width=0.8\linewidth]{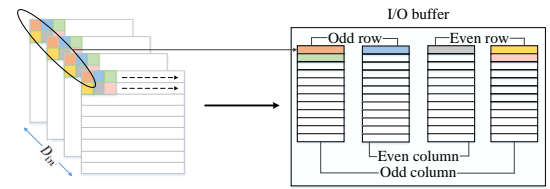}}
    \caption{Input feature layout and input buffer organization}
\label{buffer}
\vspace{-0.5em}
\end{figure}

\textbf{Input Feature Layout}
To make best use of the entire computing array, the neighboring input features must be fed to the computing array continuously. However, when the input features are sequentially stored in a single port buffer, it can not load the four features in different rows and columns of the input features from the on-chip buffer in a single cycle simply with wider read port. A four-port on-chip buffer can meet the computing requirement, but it is extremely resource-consuming in terms of both power and chip area. To address the problem, we modify both the input feature layout and the structure of the input buffer as shown in Figure \ref{buffer}. Since the four features for each BLI processing of a deformed output feature are located in adjacent rows and columns, we separate the input features into four partitions based on the feature coordinate parity in the feature map and the four features will always be located in different partitions. Accordingly, we have the buffer divided into four banks and each bank accommodate an input feature partition. Four features required by the BLI processing of any feature can be loaded in a single cycle. In addition, we have the input features stored in channel-major order and widen the port of the buffer bank such that features of multiple channels are read at the same time for all the different PE clusters. 

\begin{figure}[tb]
	\center{\includegraphics[width=0.95\linewidth]{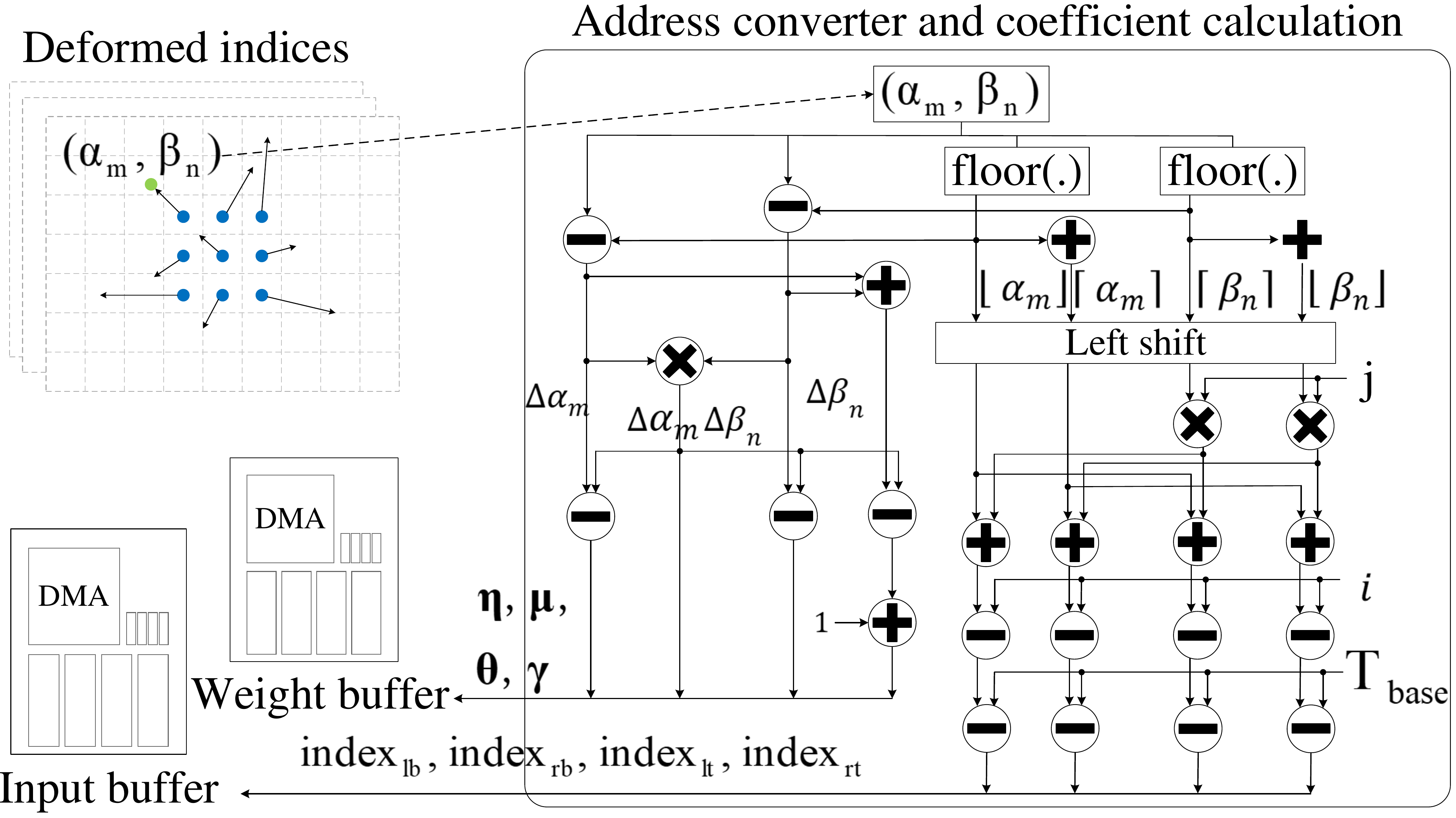}}
    \caption{Address Converter and Coefficient Calculation Block}
\label{index}
\vspace{-0.5em}
\end{figure}

\textbf{Address Converter:}
In order to fetch neighboring features for the BLI, we need to calculate the buffer addresses of the four input features based on the non-integer indices. The basic idea is to obtain the four neighboring integers of the non-integer feature indices first and then deduct the base index of the four features to calculate the on-chip buffer addresses of the required features. Essentially, it is a conversion from 3-D feature map indices to the 1-D on-chip buffer indices and the higher dimension indices including the channel index and the height index need to be scaled accordingly as formulated in Equation \ref{eq:indices}. $\alpha_{m}$ and $\beta_{n}$ denote the original non-integer feature indices i.e. coordinates in the 2D feature plane. $index_{lb}$, $index_{rb}$, $index_{lt}$, and $index_{rt}$ denote the four buffer addresses of the features located at the left bottom, right bottom, left top and right top respectively. $\mathbf{A}$ denotes the number of PEs in the computing array in the neural network accelerator, $\mathbf{H}$ denotes the height of the input feature map, $\mathbf{c}$ denotes the channel number of the input feature, $T_{0}$ denotes the base index of the four neighboring features. As the address conversion in different channels are the same, the formulation only illustrates the conversion in the 2-D feature plane. To enable runtime BLI, we have a specialized address converter module implemented. It can be easily pipelined as shown in Figure \ref{index}. The generated indices will be aligned and sent to the different input buffer banks to retrieve the corresponding four features for the BLI calculation. 

\begin{equation}
\footnotesize
\label{eq:indices}
    \left\{\begin{matrix}
index_{lb}=\left ( \left \lfloor \left \lfloor \beta _{n} \right \rfloor/2 \right \rfloor \times j+\left \lfloor \left \lfloor \alpha _{m} \right \rfloor /2\right \rfloor  \right ) \times i-T_{0}\\ 
index_{rb}=\left ( \left \lfloor \left \lfloor \beta _{n} \right \rfloor/2 \right \rfloor \times j+\left \lfloor \left \lceil \alpha _{m} \right \rceil /2\right \rfloor  \right ) \times i-T_{0}\\ 
index_{lt}=\left ( \left \lfloor \left \lceil \beta _{n} \right \rceil/2 \right \rfloor \times j+\left \lfloor \left \lfloor \alpha _{m} \right \rfloor /2\right \rfloor  \right ) \times i-T_{0}\\ 
index_{rt}=\left ( \left \lfloor \left \lceil \beta _{n} \right \rceil/2 \right \rfloor \times j+\left \lfloor \left \lceil \alpha _{m} \right \rceil /2\right \rfloor  \right ) \times i-T_{0}
\end{matrix}\right.
\; \; \; \; \; \begin{pmatrix}
i=\left \lceil c/(A/4) \right \rceil\\ 
j = H/2
\end{pmatrix}
\end{equation}

\textbf{Coefficient Calculation:} To enable runtime BLI, coefficients also need to be calculated at runtime and they are formulated in Equation~\ref{eq:bli_change} according to Equation \ref{eq:bli}. The four coefficients can be reused across the different channels, so the formulation only illustrates the calculation in the 2-D feature plane. We notice that the multiplication result $\Delta \alpha _{m}\Delta \beta _{n}$ is required by all the four coefficient calculation, so it is calculated first. Then the rest of the coefficient calculation can be conducted with only addition and subtraction. The pipelined architecture is shown on the left of \autoref{index}. The coefficients will be stored in the weight buffer for the BLI calculation according to the BLI mapping.

\begin{equation}
\footnotesize
\label{eq:bli_change}
\begin{aligned}
\left\{\begin{matrix}
    \eta =\left ( 1-\Delta \alpha _{m} \right )\left ( 1-\Delta \beta _{n} \right )\\ 
    \mu =\left ( 1-\Delta \alpha _{m} \right )\Delta _{n}\; \; \; \, \; \; \; \; \; \; \; \\ 
    \theta =\Delta \alpha _{m}\left ( 1-\Delta \beta _{n} \right )\; \; \; \; \; \; \; \; \\ 
    \gamma =\Delta \alpha _{m}\Delta \beta _{n}\: \: \: \: \: \: \: \: \: \: \: \: \: \: \: \; \; \; \; \; \, 
\end{matrix}\right.
\end{aligned}
\end{equation}

\subsection{Runtime Tile Scheduling}
In order to address the dynamic and irregular memory access problem in deformable convolution, we propose to track the data dependency at runtime and optimize the execution with runtime scheduling based on the tracked data dependency. The dependency tracking and scheduling optimization will be illustrated in the rest of this sub section.

\begin{figure}[tb]
	\center{\includegraphics[width=0.95\linewidth]{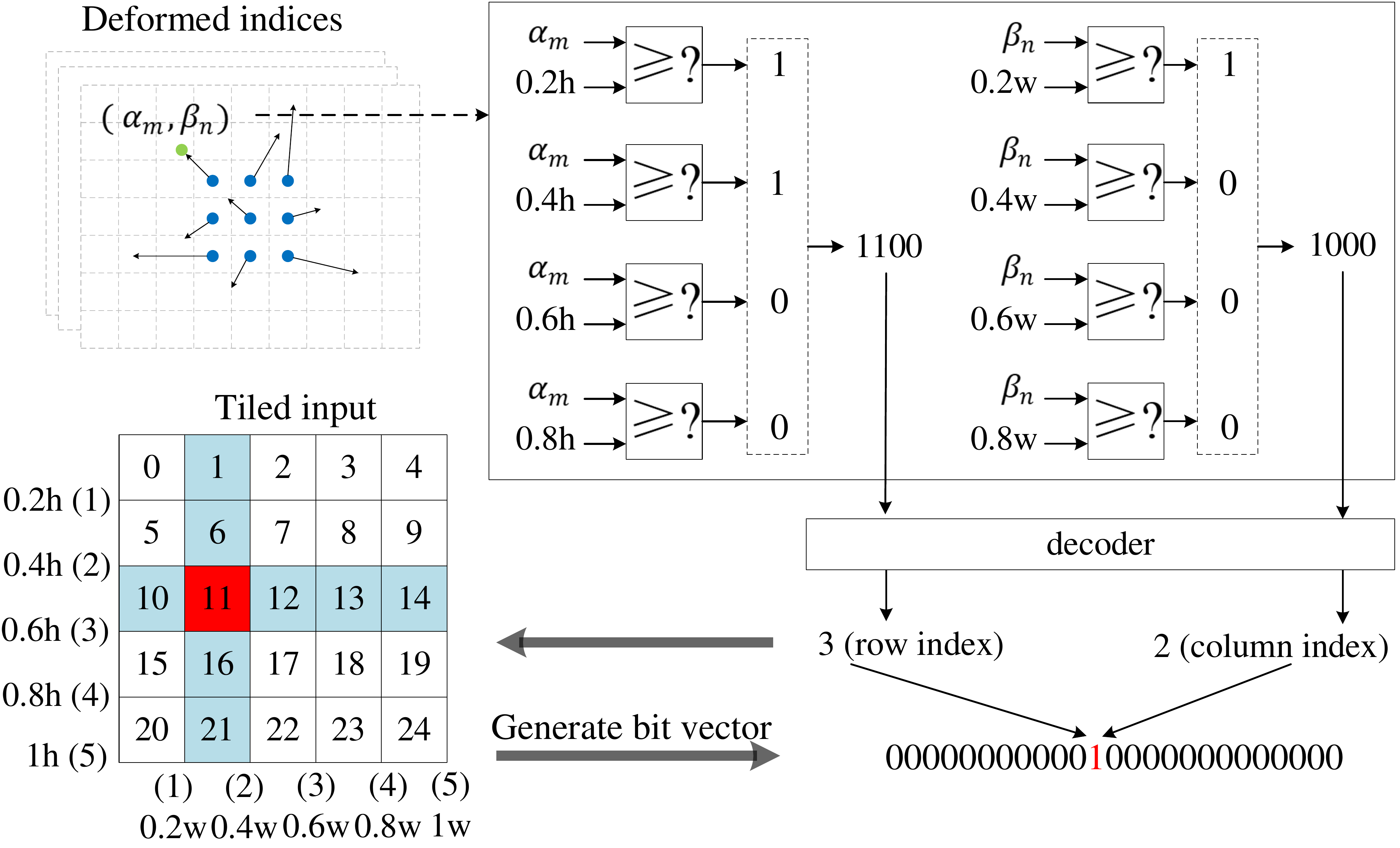}}
    \caption{Tile Dependency Table Update}
\label{fig:index2block}
\vspace{-0.5em}
\end{figure}

\textbf{Tile Dependency Tracking:} To track the dynamic memory accesses, we need a dependency table to record all the required input features for each deformable convolution output feature. Due to the limited on-chip buffer in the accelerator, the neural network processing is usually tiled and the dependency table is constructed in the granularity of a tile accordingly. The tile dependency table is abbreviated as TDT. The dependency of the deformable convolution output tile to the input feature tile can be inspected with the deformed feature indices as described in Figure \ref{fig:index2block}. Assume both the input and output features are divided into fixed $5 \times 5$ tiles. The feature indices i.e. $\alpha_{m}$ and $\beta_{n}$ are compared to the different boundary indices and the comparison result vectors can be used to determine the row index and the column index of the tile. As shown in Figure \ref{fig:index2block}, the comparison result vector for $\alpha_{m}$ is (1,1,0,0) and it means $\alpha_{m}$ is between 0.4H and 0.6H. With a decoder, we can obtain the row index of the dependent input tile. In this example, the row index is 2. Similarly, we can also obtain the column index of the dependent input tile and it is 1 in this example. Given the row index and the column index, we can decide the dependent input tile index and it is 11 in this example as highlighted with red color. With this index, we can further determine the dependent bit vector of which each bit refers to the dependency of the indexed input tile. If the corresponding input tile is dependent, it will be set to be 1 while the rest bits are set to be 0. By continuously inspecting all the deformed features required by an output tile and conducting OR operation with the bit vectors, we can obtain the entire tile dependency vector of an output tile. The tile dependency table is constructed right after the first convolution of the deformable convolution.

\textbf{Tile Scheduling:} Input tiles loaded to the on-chip buffers can be reused among the different output tile calculation, but the ordering of the input tile loading and output tile execution can both affect the reuse especially when the input buffer is limited and some of the input tiles have to be replaced. The input tile utilization varies greatly during the DCN processing as observed in the motivation section, which further aggravates the influence of the ordering of the input tile loading and the output tile execution. To address the above problem, we propose a unified tile scheduling algorithm that handles both the output tile scheduling and input tile scheduling based on the TDT that is updated at runtime. 

The proposed scheduling algorithm is presented in Algorithm \ref{alg:bitmap_sch}. It includes an output tile scheduling procedure and an input tile scheduling procedure based on the output tile scheduling result. For the output tile scheduling, it essentially selects an output tile that can reuse the input tiles that are already loaded and stored in the on-chip buffer. When the on-chip buffer is empty, it simply selects the output tile that requires the most input tiles which are more likely to be reused. When the output tile is selected, we need to determine the loading order of the dependent input tiles of the selected output tile. Although it is possible to sort all the dependent input tiles based on the its potential reuse, but the input tile reuse is expensive to estimate at runtime. In this work, we have the dependent input tiles divided into three parts as illustrated in the $input\_tile\_scheduling(.)$ function. The first part is the input tiles that are already stored in on-chip buffers. They will be scheduled first to ensure the on-chip data reuse. It can be determined by comparing the input tile on-chip status bit vector $OC$ and the dependency bit vector $B[nextID]$. The second part is the tiles that will be reused by the next output tile calculation. They will be loaded at last such that they can reside in the on-chip buffer for reuse. The next output tile calculation is obtained with the procedure $output\_tile\_scheudling(.)$. By comparing the current output tile dependency bit vector and the next output tile dependency bit vector, we can determine the overlapped input tiles that can be reused, but the tiles that are already stored in the on-chip buffer will be removed. The rest of the input tiles will be scheduled between the first part and the second part. As the input tile reuse are already considered in the scheduling, An FIFO (first in first out) strategy is used for the input tile replacement for efficient hardware implementation. 

\begin{algorithm}
    \caption{Bit vector based tile scheduling} \label{alg:bitmap_sch}
    \footnotesize
   \textbf{Input:} \\
   $N$: the number of tiles per feature map.\\ 
   $B[N][N]$: bit vector of the output tile dependency, each vector $B[i]$ ($1 \leq i \leq N$) denotes if it is dependent on the corresponding input tiles.\\
   $M$: the total number of tiles that can be stored in the on-chip buffer.\\
   $OS[N]$: bit vector of the output tiles, each bit denotes if a tile is executed.\\
   $OC[N]$: bit vector of the input tiles, each bit denotes if a tile is in the on-chip buffer.\\    
    \textbf{Output:} \\
     $OID[N]$: output tile IDs of the execution order.\\
	 $IID[N][N]$: input tile IDs of the loading order for each output tile.\\
	 
	\begin{algorithmic} [1] 
	    \State $it \gets 1$
	    \State $OID[it] \gets $ ID of the output tile that requires the most input tiles.
	    \While{$OS \neq [0,0,...,0]$}
	    \State $it \gets it + 1$
	    \State $OID[it] \gets output\_tile\_scheduling(OS, OC)$
	    \State $IID[it] \gets input\_tile\_scheduling(OID, it, OS, OC)$
	    \State $OS[OID[it-1]] \gets 0$
	    \EndWhile
	    
	    \State
	    \Procedure{$output\_tile\_scheduling$}{$OS, it$}
	    \State $TR[N]$: \# of reused tiles when an output tile is scheduled.
	    \State $currID \gets ODI[it - 1]$
	    \For{$i=1 \to N$}
	    \If{$OS[i] \neq 0$}
	    \State $vec \gets B[currID] \& B[i]$
	    \State $TR[i] \gets$ number of bit '1' in $vec$
	    \Else
	    \State $TR[i] \gets 0$
	    \EndIf
	    \EndFor
	    \State find output tile ID $ID$ that $TR[i]=max(TR)$.
	    \State \textbf{return} $ID$
	    \EndProcedure
	    
	    \State
	    \Procedure{$input\_tile\_scheduling$}{$OID, it, OS, OC$}
	    \State Suppose the input tile IDs to be loaded for next output tile is $tmpID[N]$.
	    \State $currID \gets OID[it - 1]$
	    \State $nextID \gets OID[it]$
	    \State $loadedVec \gets OC \& B[nextID]$
	    \State $lastLoadVec \gets B[currID] \& B[nextID] \& (!loadedVec)$
	    \State $seqLoadVec \gets B[nextID] \& (!loadedVec) \& (!lastLoadVec)$
	    \State push the non-zero IDs of $loadedVec$ to $tmpID$.
	    \State push the non-zero IDs of $seqLoadVec$ to $tmpID$.
	    \State push the non-zeros IDs of $lastLoadVec$ to $tmpID$.
	    \State \textbf{return} $tmpID$.
	    \EndProcedure

	\end{algorithmic}
\end{algorithm}

The proposed tile scheduling is implemented with customized hardware rather than a software scheduling algorithm on CPUs to ensure efficient tile-based execution. The tile scheduling module is shown in Figure \ref{fig:index2block}. It mainly depends on the TDT to select the next output tile for execution on the computing array in the accelerator. The basic idea is to choose the output tile that has the most dependent input tiles overlapped with that of the scheduled current output tile. Thus, we have the dependency bit vector of current output tile $AND$ with the bit vectors of all the un-executed output tiles. The $AND$ result will be sent to an non-zero (NZ) bit counter module that mainly consists of an adder tree to count the number of the non-zero bits. The number will pass through a pipelined comparator to determine the maximum value. The corresponding output tile has the most input tiles overlapped with that of the current output tile, so it will be scheduled for the execution next. Instead of having the output tile scheduling and the execution conducted sequentially, we adopt a pre-scheduling strategy that performs the next output tile scheduling in parallel with the current output tile execution. Since the execution does not have to wait for the immediate scheduling result, more complex scheduling algorithm can be implemented. When the next output tile is selected, we will schedule the dependent input tiles. The input tile scheduling mainly depends on three hardware-friendly bit-wise operations which divide the input tiles into three parts as already discussed in Algorithm \ref{alg:bitmap_sch}. By inspecting the non-zero bit number of the three resulting bit vectors with corresponding NZ bit counters, we can determine the $IDs$ of the input tiles in each partition and push them into three independent queues. As each queue belongs to different scheduling priorities, they can be scheduled sequentially and the input tile scheduling is completed when all the queues are empty.

\begin{figure}[tb]
	\center{\includegraphics[width=1\linewidth]{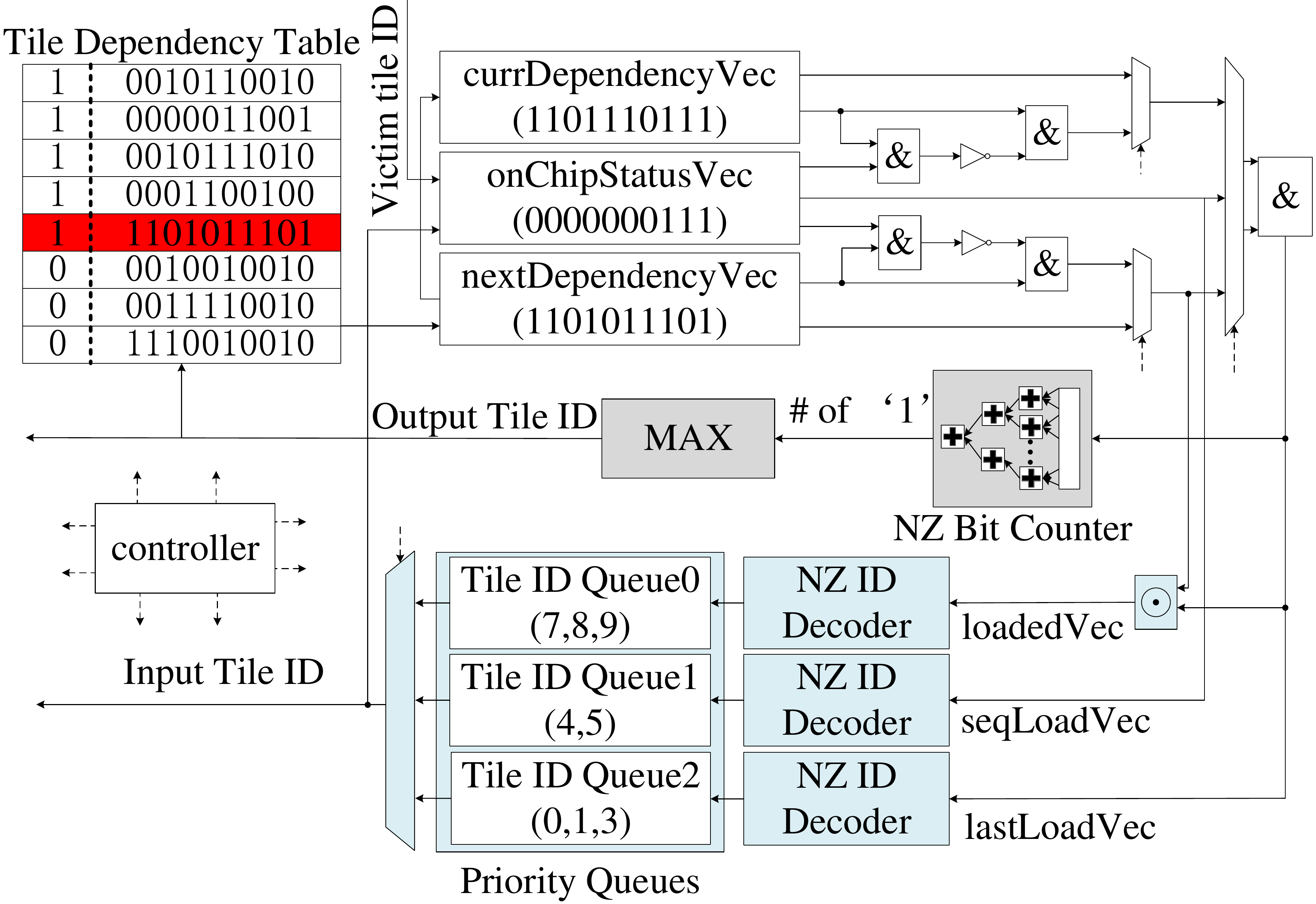}}
    \caption{Bit vector based runtime tile scheduling. The grey blocks are mainly used for the output tile scheduling while the light blue blocks are mainly used for the input tile scheduling. The rest of the blocks are shared by both the output tile scheduling and the input tile scheduling.}
\label{fig:b-compare}
\vspace{-0.5em}
\end{figure}

\subsection{BLI and Convolution Fusion}
We notice that massive data movement is required when the different processing stages of the deformable convolution is performed sequentially due to the limited on-chip buffer. Inspired by the neural network fusion techniques \cite{alwani2016fused}, we try to fuse the different processing stages such that the intermediate data can be reused via on-chip buffers without additional external memory accesses. Basically, we have the input data of the upstream processing stage tiled. When a tile of the output data is obtained in the upstream processing stage, they will be used by the downstream processing stage immediately by simply swapping the input buffer and output buffer. Compared to the stage by stage processing, the fused processing on top of the tiling avoids transferring the intermediate data to and back from the external memory, which is beneficial to both the performance and energy efficiency. While the amount of the deformed indices is usually small compared to the feature data and can be buffered on-chip directly, we mainly have the second processing stage and the third processing stage tiled and fused in practice. 
\section{Experiment} \label{sec:result}
\subsection{Experiment setup}
\textbf{Hardware Platforms:} The proposed DCN accelerator is implemented with Verilog and synthesized with Synopsys Design Compiler under TSMC 40nm library. It works at 800 MHz. The configurations of the DCN accelerator are shown in Table \ref{Accelerator parameters}. Processing elements in the accelerator adopt 8bit fixed point. We also have DCN implemented on four typical architectures including an ARM processor (ARM), ARM+TPU, GPU, and DCN Accelerator (DCNA) for comparison. The ARM processor is ARM-A7@900MHz equipped with 1GB DRAM (DDR3) which is the core of Raspberry Pi 3. The GPU is 256-core NVIDIA Pascal GPU with 8GB GDDR5 memory and it is the core of Nvidia TX2. Experiments on the ARM processor and the GPU were implemented with PyTorch 1.3 on real platforms i.e. Raspberry Pi and TX2 respectively. Experiments for ARM+TPU were conducted in a mixed manner. The second stage of the deformable convolution is not supported by TPU and it was performed on the ARM processor instead, while the rest of the neural networks was performed on TPU and evaluated with Scale-Sim \cite{samajdar2018scale}. The configurations of the TPU architecture are the same with that used in the DCN accelerator. In addition, both the ARM processor and TPU have 1GB DRAM equipped. The average power of the ARM processor is 1.3W and its idle power is 0.3W. In order to evaluate the power consumption of DRAM, we accumulate the power consumption of the different memory operations such as Activation (ACT), Read (RD), Write (WR), and Background (BG) power based on Table \ref{DDR power parameters} according to Micron's power calculators\cite{DRAM}.

\begin{table}
    \centering
  \caption{Accelerator parameters}
  \label{Accelerator parameters}
  \begin{tabular}{ccccccc}
    \toprule  
      \# of PEs & In Buf & Out Buf & Weight Buf & Index Buf & Inst Buf \\
    \midrule
      16$\times$32 & 128KB & 256KB & 256KB & 32KB & 64KB\\
  \bottomrule
\end{tabular}
\vspace{-0.1em}
\end{table}

\begin{table}
    \centering
  \caption{Power consumption of each different memory operations}
  \label{DDR power parameters}
  \begin{tabular}{cccccc}
    \toprule
      ACT & RD & WR & READ I/O & Write ODT & BG \\
    \midrule
      63.7mW & 52.1 mW & 52.1 mW & 32.7 mW & 136.1mW & 67.7 mW \\
  \bottomrule
\end{tabular}
\vspace{-0.1em}
\end{table}

\textbf{Neural Network Benchmark:} In order to evaluate the proposed DCNA, we have two typical neural network models including VGG19\cite{simonyan2014very} and SegNet\cite{badrinarayanan2017segnet} used as our benchmark. Deformable convolution can be used to replace any convolution in neural networks, but the replacement configurations can lead to different trade-offs between computation and model accuracy. In this case, we have three typical deformable convolution configurations set for each model and they are denoted as VGG19/SegNet-3, VGG19/SegNet-8, and VGG19/SegNet-F. As the convolution layers close to the output layer are usually smaller, we have deformable convolution placed from the output layer to input layer of the neural networks to minimize the deformable convolution induced computation. VGG19/SegNet-3 and VGG19/SegNet-8 represents that the last three convolution layers and the last eight convolution layers of VGG19/SegNet are replaced with the deformable convolution layers respectively. VGG19/SegNet-F represents that all the convolution layers are replaced with deformable convolution layers. Details of the benchmarks are summarized in Table~\ref{benchmark}.

\begin{table}
    \centering
  \caption{Neural Network Benchmark}
  \label{benchmark}
  \begin{tabular}{ccccc}
    \toprule  
      Network & \# of deformable Conv & \# of Conv & Kernel types\\
    \midrule
      VGG19-3 & 3 & 13 & 3\\
      VGG19-8 & 8 & 8 & 3\\
      VGG19-F & 19 & 0 & 3\\
      SegNet-3 & 3 & 13 & 3\\
      SegNet-8 & 8 & 8 & 3\\
      SegNet-F & 16 & 0 & 3\\
  \bottomrule
\end{tabular}
\vspace{-1em}
\end{table}

\subsection{Performance Evaluation}
\begin{figure}
	\center{\includegraphics[width=0.9\linewidth]{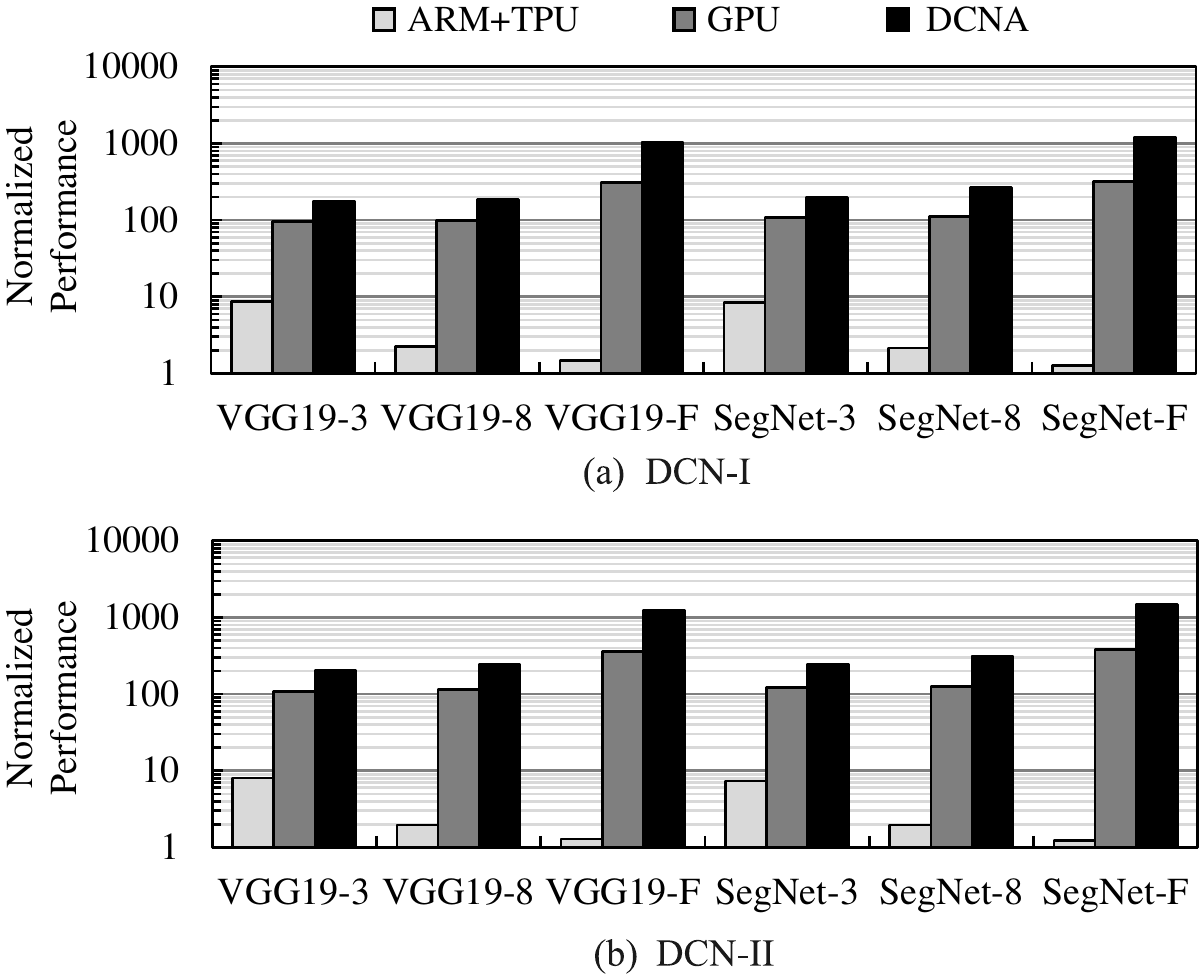}}
    \caption{Normalized (to ARM) performance of DCNs on different computing architectures.}
\label{performance}
\vspace{-0.5em}
\end{figure}

The performance of the DCN execution on the different computing architectures is normalized to that on an ARM processor. As shown in Figure \ref{performance}, DCNA achieves 515$\times$ and 621$\times$ higher performance on DCN-I and DCN-II respectively on average compared to a general ARM processor. DCN-II requires more sampling locations in deformable convolution operations of DCNs, so it has more computation and random accesses involved, which leads to larger execution time. Accordingly, higher performance speedup is achieved for DCN-II on DCNA. On the other hand, we notice that DCNA achieves much higher performance speedup on VGG19/SegNet-F compared to the other two configurations (i.e. VGG19/SegNet-3 and VGG19/SegNet-8) with less deformable convolution operations in the neural networks. The main reason is that the deformable convolution is rather challenging for the ARM processor due to the irregular memory accesses. The execution of deformable convolution operations on an ARM processor dominates the total DCN execution time and becomes the performance bottleneck of DCNs. In contrast, convolution with regular memory accesses is usually intensively optimized on ARM processors in PyTorch and the performance speedup achieved on customized accelerators is relatively lower. When DCNs are deployed on the ARM+TPU architecture, we can only have the convolution operations accelerated with TPU while the deformable convolution is still executed on the ARM processor. According to Amdahl's law, the deformable convolution remains the performance bottleneck. Thereby, the performance speedup of the ARM+TPU architecture to the ARM processor is rather low especially for VGG19/SegNet-F. Unlike the ARM+TPU architecture, GPU in TX2 can have the entire DCNs implemented and the deformable convolution operations can also benefit from the GPU parallel processing due to its powerful support on general tensor operations. Thus, significant performance speedup is achieved compared to the ARM+TPU architecture. DCNA that is built on top of a conventional neural network accelerator has customized circuit design for both the standard convolution and the new deformable convolution outperforms GPU and exhibits 2.21$\times$ performance speedup on average.

\subsection{Energy Consumption Evaluation}
\begin{figure}
	\center{\includegraphics[width=0.95\linewidth]{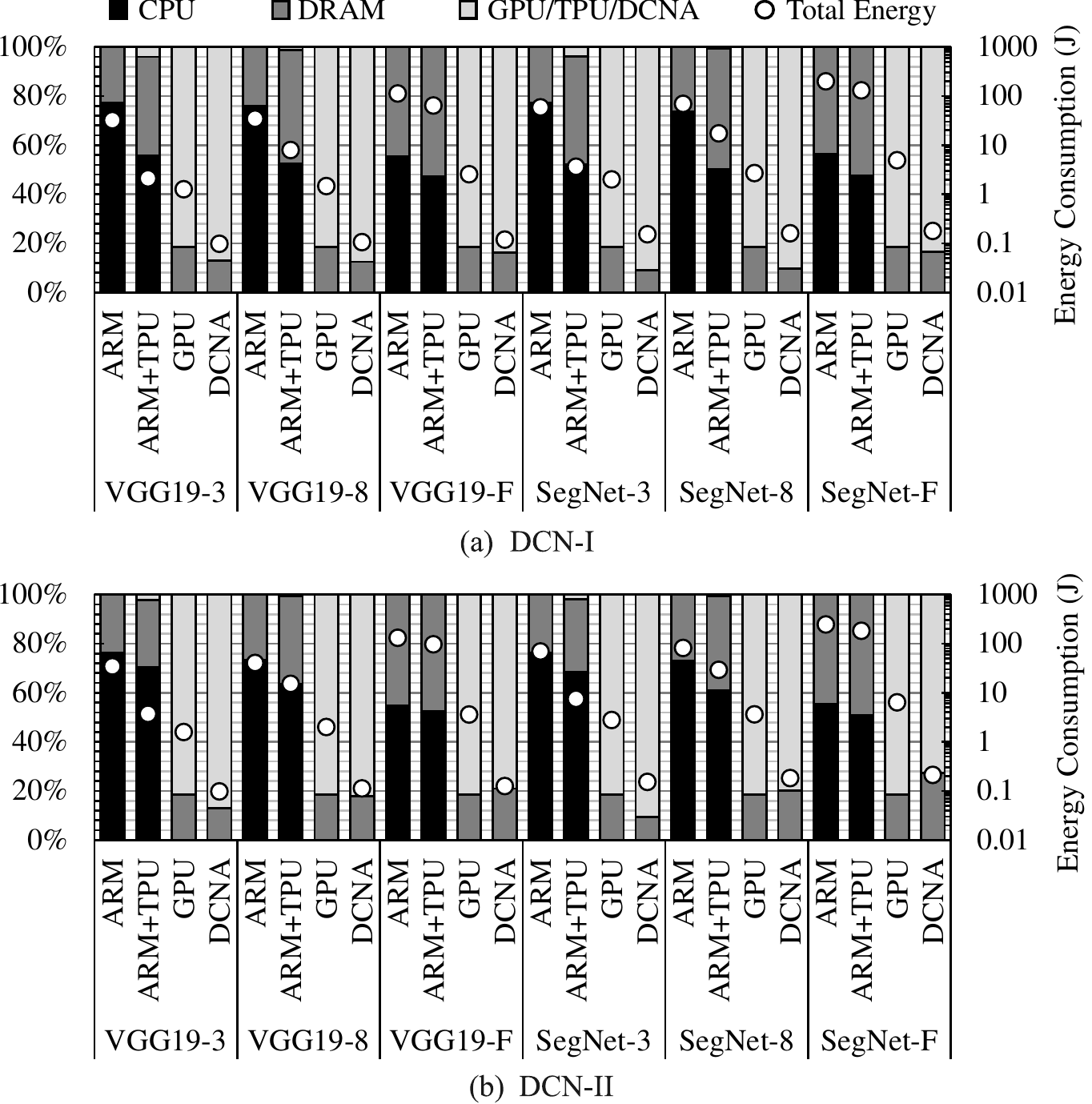}}
    \caption{Energy consumption of DCNs on four computing architectures}
\label{energy}
\vspace{-0.5em}
\end{figure}

The energy consumption of DCNs on the four computing architectures is presented in \autoref{energy}. It shows both the total energy consumption and the energy consumption distribution. Generally, DCNA with customized hardware acceleration shows the lowest energy and it is 612$\times$ less than the ARM processor. The energy consumption benefit is attributed to both the much smaller execution time brought by the performance acceleration as illustrated in \autoref{performance} and the lower power consumption. For the ARM+TPU architecture, although TPU can accelerate the convolution operations and consumes little energy, it still consumes considerable time and energy for the deformable convolution operations on the ARM processor. GPU on Nvidia Jetson TX2 can have both the convolution and deformable convolution accelerated, so it greatly reduces the execution time but its energy consumption remains $9 \times$ higher on average than DCNA due to the much higher power consumption. Moreover, we notice that the percentage of the DRAM energy consumption on VGG19/SegNet-F is larger than that on VGG19/SegNet-3 on DCNA. The main reason is that VGG19/SegNet-F with more deformable convolution involves many irregular memory accesses and lowers the memory access efficiency. Particularly, DCNA handles the irregular memory accesses with the granularity of a tile. Usually only a portion of the data in a tile is required and many of the data remain unused though the dependency is considered by the on-chip tile scheduler. In addition, some of the tiles may have to be repeatedly loaded due to the limited on-chip buffer and irregular data reuse. Thereby, the memory access efficiency is much lower than that of a standard convolution with regular memory access patterns. The lower memory accesses eventually lead to higher DRAM energy consumption.


\subsection{Chip Area Evaluation}
\begin{figure}
	\center{\includegraphics[width=0.95\linewidth]{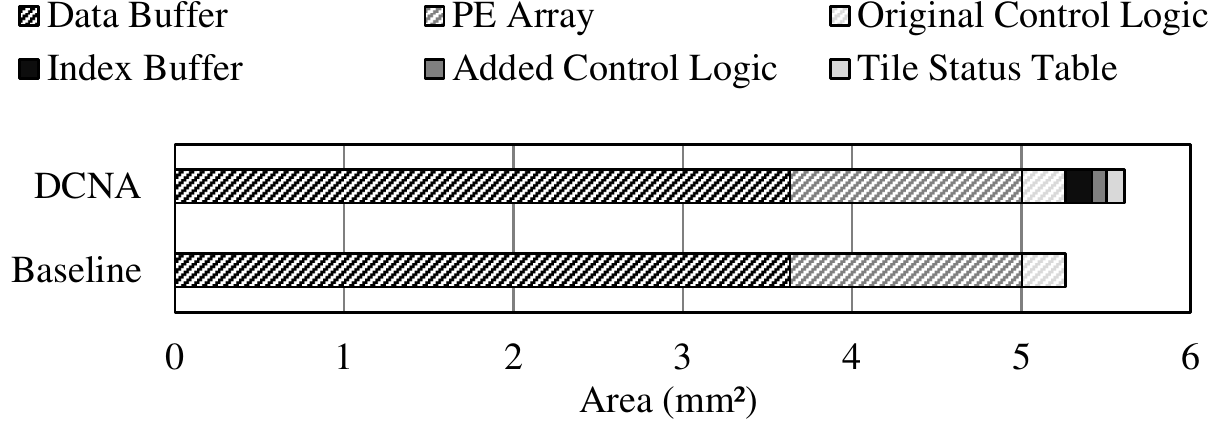}}
    \caption{Chip area of the baseline neural network accelerator and DCNA}
\label{area}
\vspace{-0.5em}
\end{figure}

The baseline neural network accelerator can be roughly divided into on-chip Data Buffer (input feature/ output feature/ weight/ bias buffer), PE Array, and the Original Control Logic while DCNA requires additional components including Index Buffer, Tile Dependency Table, and the Added Control Logic. The chip area of the different components is presented in \autoref{area}. It can be seen that DCNA induces only 6.6\% chip area compared to the baseline design. Among the added hardware blocks, index buffer that usually needs to store the generated feature indices in a deformable convolution operation takes up the most chip area. In contrast, the additional control logic and the tile dependency table consume negligible chip area. Since the feature indices generated in deformable convolution is usually reused among the different channels, it is much less than the input/output features and weights. Hence, the index buffer size is much smaller compared to the data buffer in the baseline neural network accelerator.

\subsection{Optimization Evaluation}
\begin{figure}
	\center{\includegraphics[width=0.90\linewidth]{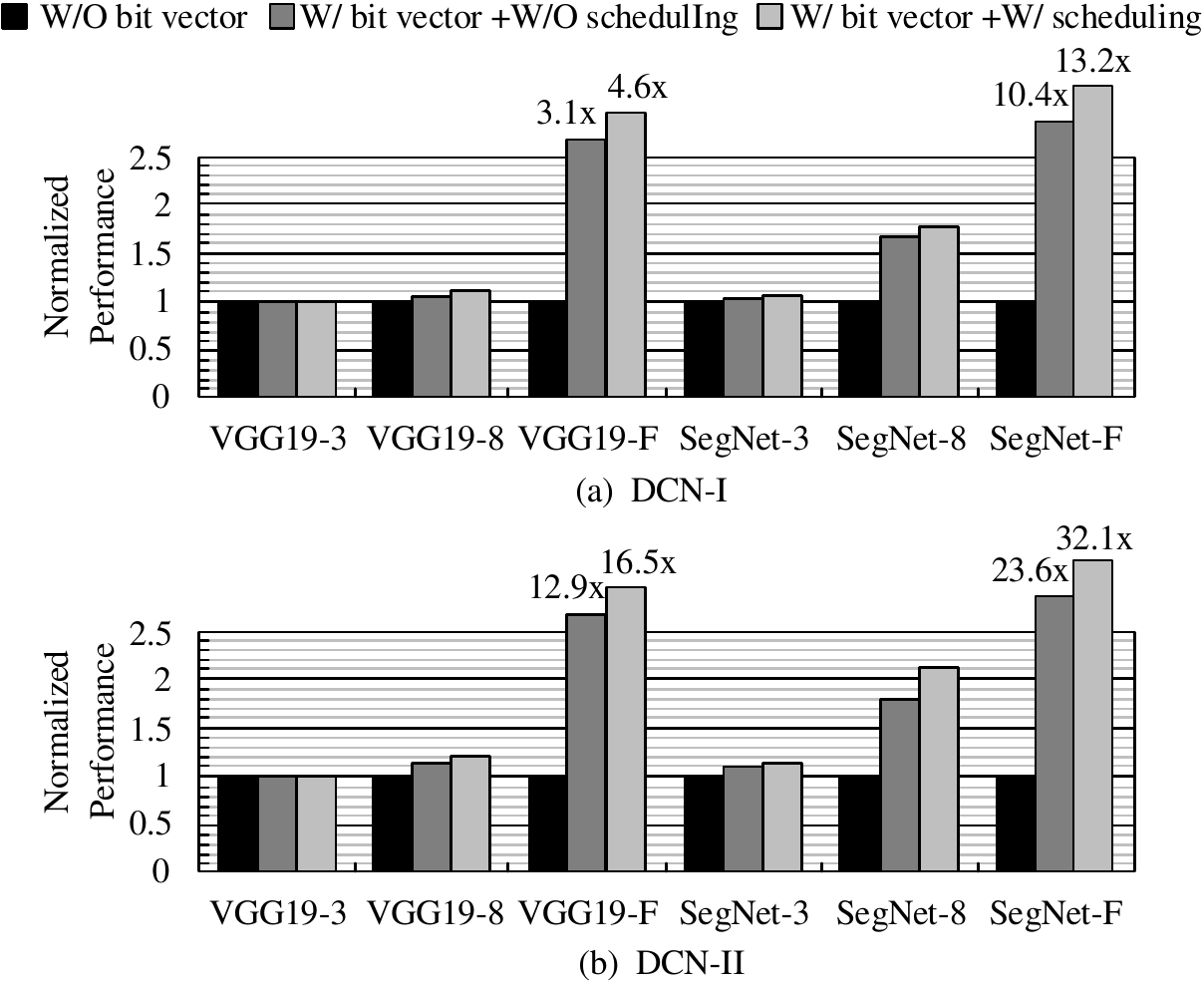}}
    \caption{Influence of tile scheduling on the DCN performance.}
\label{bitmap_performance}
\vspace{-0.5em}
\end{figure}

\textbf{Tile Scheduling:} To evaluate the influence of the proposed tile scheduling on the DCN performance, we compare it with a naive implementation without bit vector based dependency tracking (W/O bit vector) and an implementation with bit vector based tracking but without tile scheduling (W/ bit vector + W/O scheduling). The naive implementation executes the output features sequentially and loads all the dependent input tiles as needed due to the lack of the overall dependency information. The implementation with bit vector based tracking but without tile scheduling sequentially executes all the output tiles, but it loads the dependent input tiles of the entire output tile rather than a single output feature based on the corresponding TDT. The loaded input tiles will also be used for the calculation of all the features in the output tile such that the loaded tiles can be reused as much as possible. The proposed implementation has both the bit vector based tracking and tile scheduling considered (W/ bit vector + W/ scheduling). It optimizes the ordering of both the output tile execution and the input tile loading for more efficient data reuse. The comparison is shown in \autoref{bitmap_performance}. It can be seen that VGG19/SegNet-F with all the convolution layers replaced with deformable convolution benefits most from the tile-based dependency tracking and scheduling. In contrast, VGG19/SegNet-8 have only three small convolution layers replaced with the deformable convolution exhibits marginal performance speedup. It is mainly caused by two reasons. First, the computation of the deformable convolution operations takes up only a small portion of the entire neural network computation, there is little space left for the performance improvement. Second, the sizes of the deformable convolution operations are small and the data including the input features, the weights, and the output features can be mostly fully buffered. Hence, the tiles can be reused without any scheduling and the proposed tile scheduling shows no performance improvement in this case. The performance improvement on DCN-I and DCN-II also differs. This is mainly caused by the fact that DCN-II involves more random sampling and thus benefits more from the DCNA acceleration. 

\begin{figure}
	\center{\includegraphics[width=0.90\linewidth]{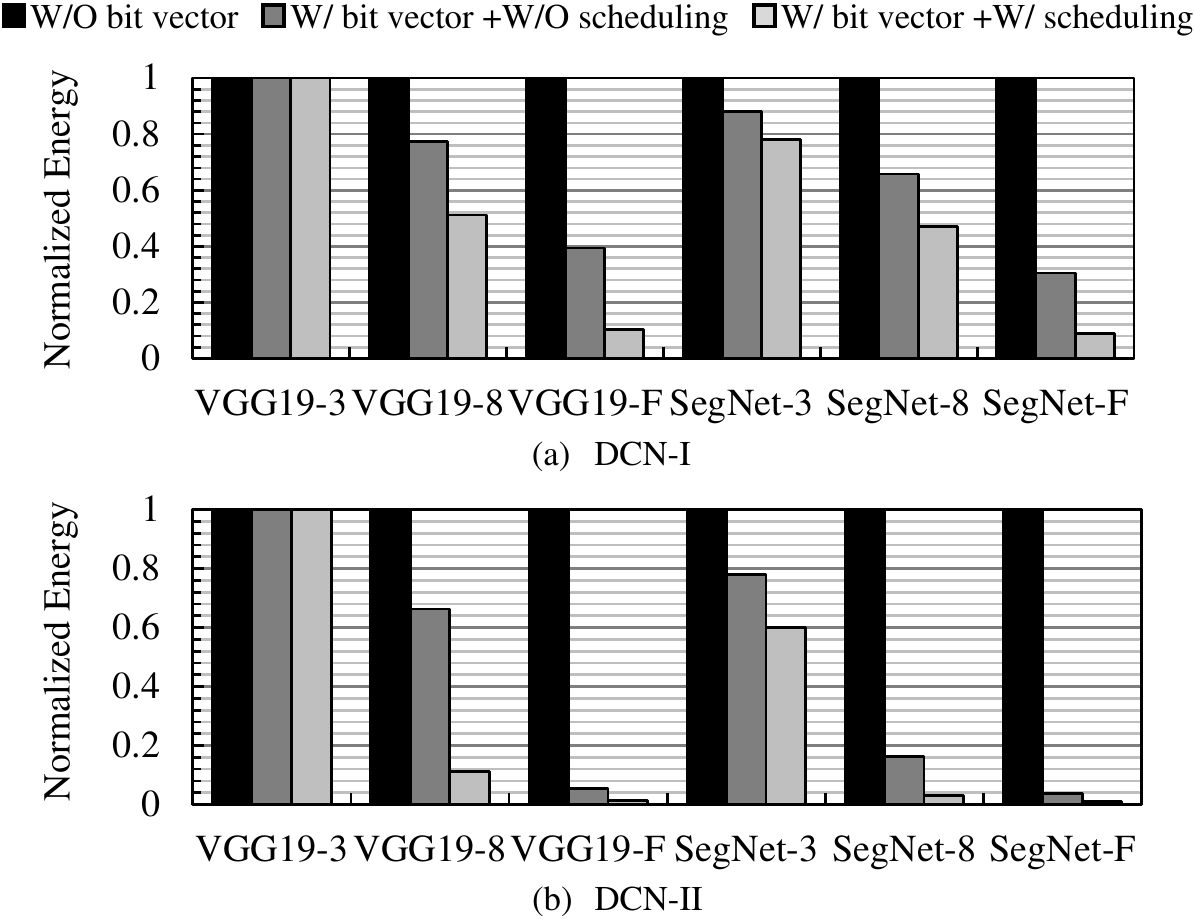}}
    \caption{Influence of tile scheduling on the DCN energy consumption.}
\label{percent_energy}
\vspace{-0.5em}
\end{figure}

\begin{figure}
	\center{\includegraphics[width=0.90\linewidth]{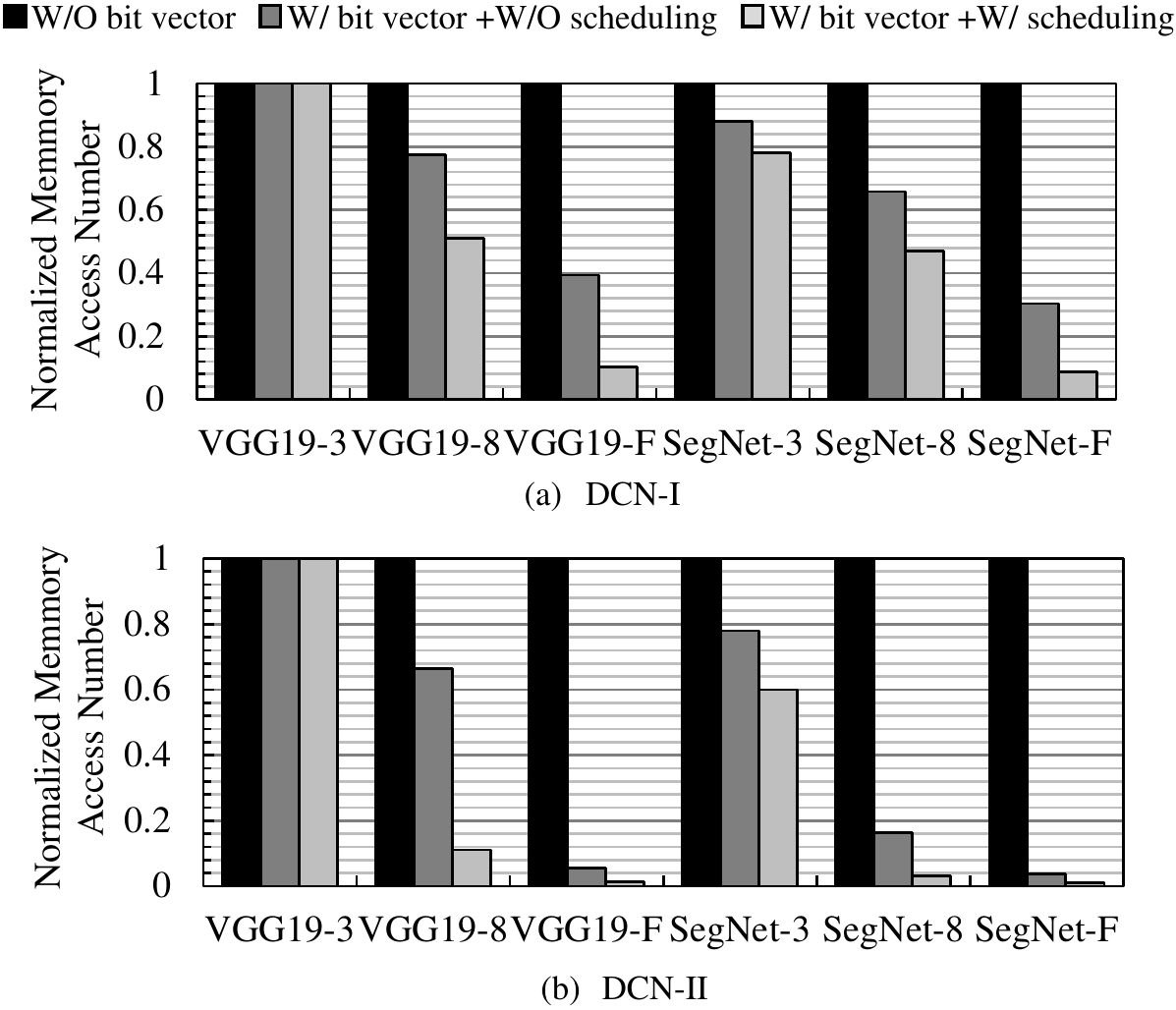}}
    \caption{Influence of tile scheduling on the number of DCN memory accesses.}
\label{mem_access}
\vspace{-0.5em}
\end{figure}

On top of the performance improvement, we also evaluate the influence of the tile scheduling on the energy consumption of the DCN execution. The experiment result is shown in \autoref{percent_energy}. Again, it can be seen that VGG19/SegNet-F with most deformable convolution operations benefits most and shows the least energy consumption while VGG19/SegNet-3 with small deformable convolution operations has little optimization space for the tile scheduling. Generally, the energy reduction is mainly attributed to both the reduced execution time according to \autoref{bitmap_performance} and the lower power consumption brought by the reduced memory accesses. As the performance improvement is already discussed in prior subsection, we mainly investigate the memory access reduction in this subsection. The total memory accesses issued by DCNA during the DCN processing is shown in \autoref{mem_access}. By comparing the implementation without bit vector based dependency tracking and the implementation with bit vector but no scheduling, we observe that the bit vector based tile dependency tracking removes substantial memory accesses. It is mainly achieved by avoiding repeatedly loading the same input tiles required by the calculation of different output features in the same output tile. The scheduling further reduces the memory accesses by inspecting the input tile reuse among the different output tiles according to the tile dependency table. The experiment shows that the proposed scheduling reduces the memory accesses by 40.7\% on VGG19/SegNet-F on average compared to the implementation with only bit vector based dependency tracking.

\begin{figure}
	\center{\includegraphics[width=0.8\linewidth]{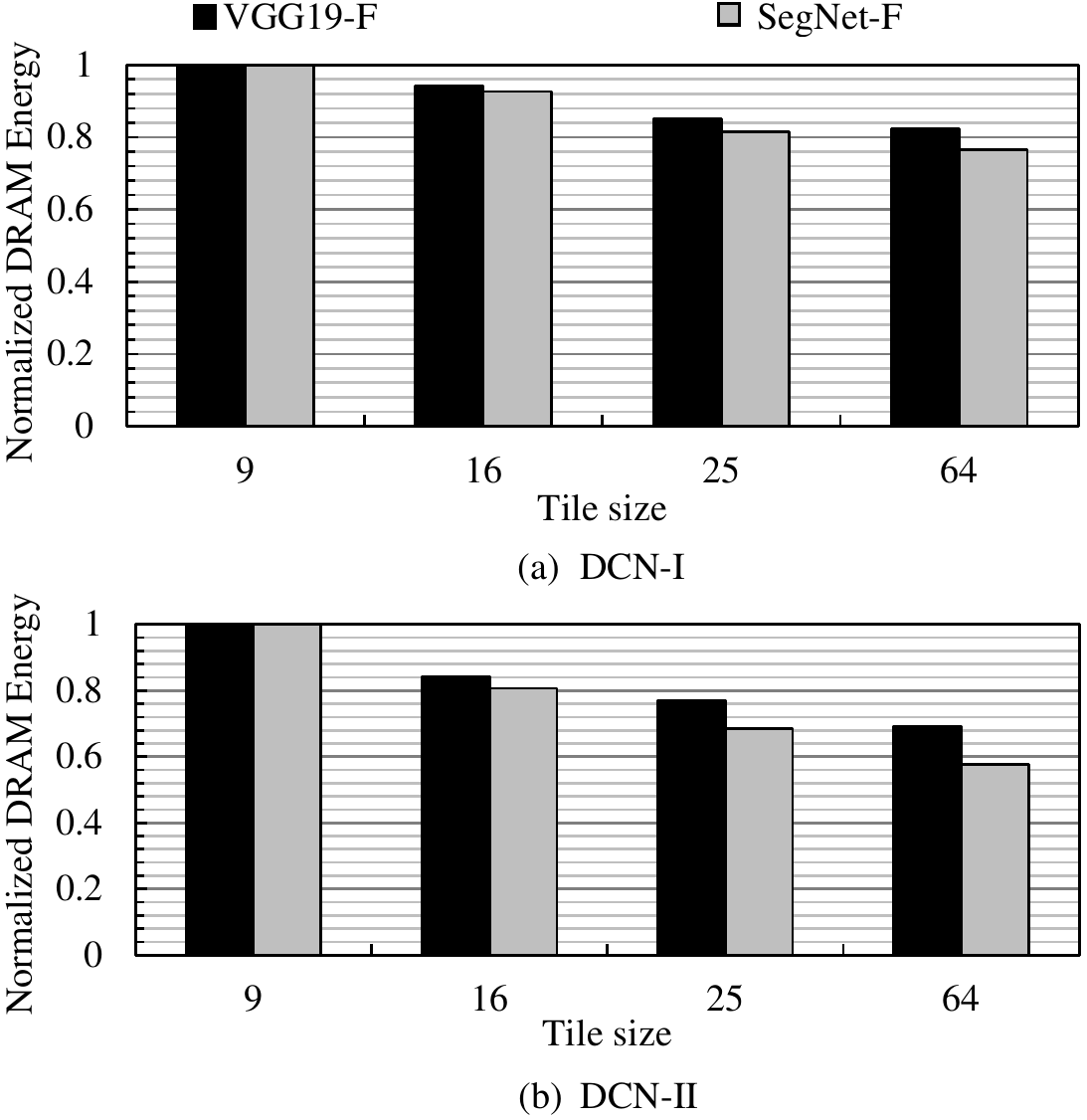}}
    \caption{Normalized DRAM energy consumption under different tile sizes.}
\label{tile_scale}
\vspace{-0.5em}
\end{figure}

\textbf{Tile Sizing}: Tile size is an important design parameter and it determines the granularity of the tile dependency tracking and scheduling. Since most of the memory access latency in DCNA can be overlapped by the computing latency, tile size mainly affects the memory access efficiency and the DRAM energy consumption eventually, and has little influence on the performance. We take VGG19/SegNet-F with most deformable convolution operations as an example and analyzed the DRAM energy consumption under different tile sizes. The experiment result is shown in \autoref{tile_scale}. It can be seen that the DCN processing with smaller tile size benefits most from DCNA and consumes the least DRAM energy. The main reason is that smaller tile size allows more efficient tile dependency tracking and scheduling. Thus, the on-chip buffer utilization and DRAM memory access efficiency can be improved. In contrast, when the tile size is large, the dependent input features of each output tile can spread across all the input tiles. In this case, all the input tiles have to be repeatedly loaded for each output tile calculation, leaving little optimization space for the proposed tile dependency tracking and scheduling. 

\begin{figure}
	\center{\includegraphics[width=0.88\linewidth]{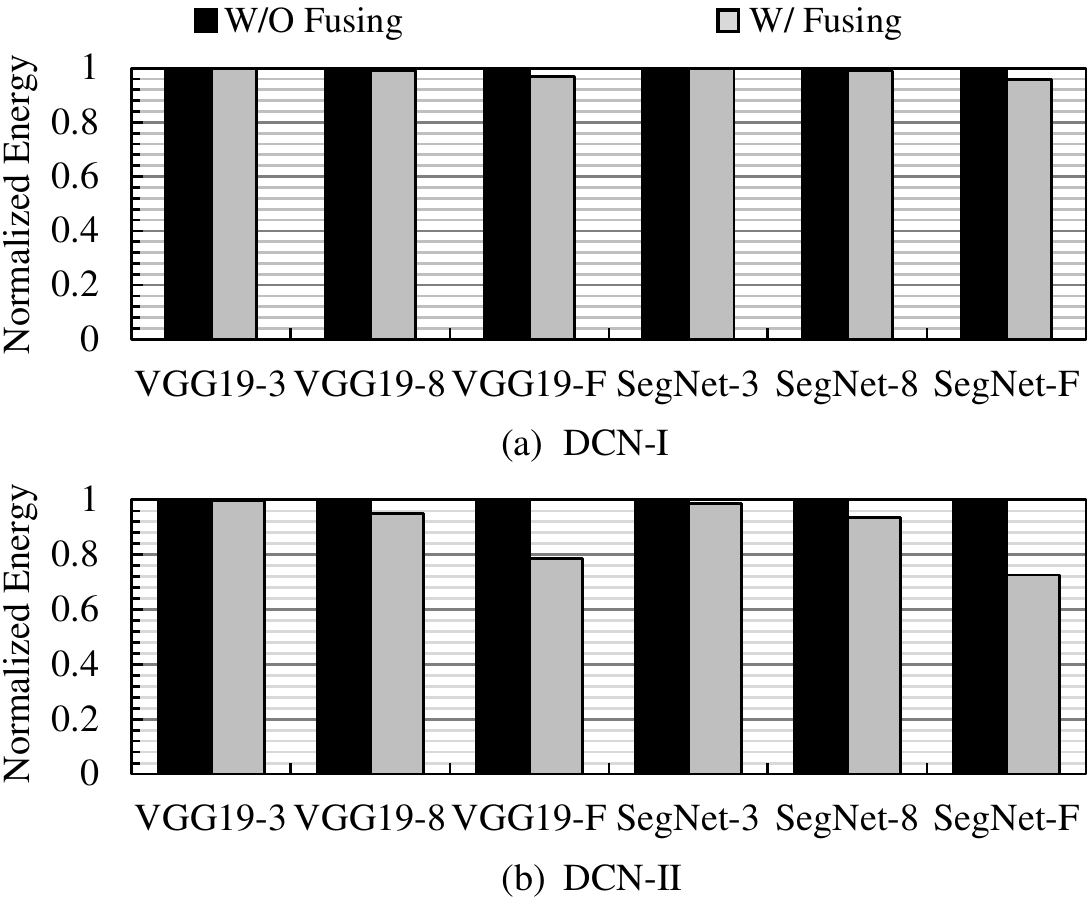}}
    \caption{Influence of layer fusion on the DCN energy consumption.}
\label{fusing}
\vspace{-0.5em}
\end{figure}

\textbf{Layer Fusion:} In order to reduce the intermediate data transmission to and from DRAM, we propose to fuse the processing stages in each deformable convolution operation. The influence of the layer fusion on the energy consumption is shown in \autoref{fusing}. It can be seen that the fusion reduces the energy consumption by more than 20\% on VGG19/SegNet-F with DCN-II structure. The main reason is that the memory access time in the two DCNs cannot be fully overlapped by the computation time due to the involved large deformable convolution operations and the fusion that avoids large intermediate data transmission improves the DCN performance other than the memory access efficiency. Unlike VGG19/SegNet-F, the deformable convolution operations in the rest scenarios are relatively small and the fusion only reduces the memory accesses while the performance has little improvement. As the DRAM energy consumption takes up only around 20\% of the entire DCN energy consumption according to \autoref{energy}, the energy reduction brought by the memory reduction is limited. Particularly, for VGG19/SegNet-3, the entire intermediate data can be mostly fully buffered and there is even little memory access optimization space left for the fusion. Thereby, the fusion shows little energy reduction.

\section{Conclusion} \label{sec:conclusion}
DCNs that have been demonstrated to be effective in many practical scenarios with image geometric or photometric variations include new deformable convolution operations other than the conventional neural network operations. The deformable convolution operations that require random sampling over the entire input feature maps incur considerable irregular memory accesses and bilinear interpolation (BLI) operations, which that can not be fitted to the existing neural network accelerator architecture. In this work, we revisit the conventional neural network architecture by introducing a runtime tile-based data dependency tracking and scheduling mechanism to address the irregular memory accesses and optimize the data reuse in DCNs. At the same time, we reorganize the BLI operations to fit them to the 2D computing array for parallel processing. Finally, we also fuse the different processing stages in each deformable convolution operation to enable on-chip data reuse and reduce the intermediate data transmission via DRAM. According to our experiments on representative neural networks with different deformable convolution configurations, the proposed DCNA that supports both the convolution and deformable convolution achieves 45$\times$-546$\times$ performance speedup over the ARM+TPU architecture that relies on the ARM processor to handle the deformable convolution. When compared to GPU that can execute the entire DCNs, DCNA shows 3$\times$ performance speedup and 18.6$\times$ energy reduction. 

\bibliographystyle{IEEEtran}
\bibliography{refs}

\begin{thebibliography}{10}
\providecommand{\url}[1]{#1}
\csname url@samestyle\endcsname
\providecommand{\newblock}{\relax}
\providecommand{\bibinfo}[2]{#2}
\providecommand{\BIBentrySTDinterwordspacing}{\spaceskip=0pt\relax}
\providecommand{\BIBentryALTinterwordstretchfactor}{4}
\providecommand{\BIBentryALTinterwordspacing}{\spaceskip=\fontdimen2\font plus
\BIBentryALTinterwordstretchfactor\fontdimen3\font minus
  \fontdimen4\font\relax}
\providecommand{\BIBforeignlanguage}[2]{{%
\expandafter\ifx\csname l@#1\endcsname\relax
\typeout{** WARNING: IEEEtran.bst: No hyphenation pattern has been}%
\typeout{** loaded for the language `#1'. Using the pattern for}%
\typeout{** the default language instead.}%
\else
\language=\csname l@#1\endcsname
\fi
#2}}
\providecommand{\BIBdecl}{\relax}
\BIBdecl

\bibitem{dai2017deformable}
J.~Dai, H.~Qi, Y.~Xiong, Y.~Li, G.~Zhang, H.~Hu, and Y.~Wei, ``{Deformable
  convolutional networks},'' in \emph{Proceedings of the IEEE international
  conference on computer vision}, 2017, pp. 764--773.

\bibitem{cao2019object}
Z.~Cao, X.~Li, and L.~Zhao, ``Object detection in vhr image using transfer
  learning with deformable convolution,'' in \emph{IGARSS 2019-2019 IEEE
  International Geoscience and Remote Sensing Symposium}.\hskip 1em plus 0.5em
  minus 0.4em\relax IEEE, 2019, pp. 326--329.

\bibitem{8953851}
C.~Zhang and J.~Kim, ``{Object Detection With Location-Aware Deformable
  Convolution and Backward Attention Filtering},'' in \emph{2019 IEEE/CVF
  Conference on Computer Vision and Pattern Recognition (CVPR)}, 2019, pp.
  9444--9453.

\bibitem{8901912}
D.~Zhang, L.~Li, Z.~Zhu, S.~Jin, W.~Gao, and C.~Li, ``{Object Detection
  Algorithm Based on Deformable Convolutional Networks for Underwater
  Images},'' in \emph{2019 2nd China Symposium on Cognitive Computing and
  Hybrid Intelligence (CCHI)}, 2019, pp. 274--279.

\bibitem{deng2019restricted}
L.~Deng, M.~Yang, H.~Li, T.~Li, B.~Hu, and C.~Wang, ``Restricted deformable
  convolution-based road scene semantic segmentation using surround view
  cameras,'' \emph{IEEE Transactions on Intelligent Transportation Systems},
  2019.

\bibitem{chen2018encoder}
L.-C. Chen, Y.~Zhu, G.~Papandreou, F.~Schroff, and H.~Adam, ``Encoder-decoder
  with atrous separable convolution for semantic image segmentation,'' in
  \emph{Proceedings of the European conference on computer vision (ECCV)},
  2018, pp. 801--818.

\bibitem{8953750}
Y.~Xiong, R.~Liao, H.~Zhao, R.~Hu, M.~Bai, E.~Yumer, and R.~Urtasun, ``{UPSNet:
  A Unified Panoptic Segmentation Network},'' in \emph{2019 IEEE/CVF Conference
  on Computer Vision and Pattern Recognition (CVPR)}, 2019, pp. 8810--8818.

\bibitem{diao2020multi}
Y.~Diao, J.~Chen, and Y.~Qian, ``{Multi-Label Remote Sensing Image
  Classification with Deformable Convolutions and Graph Neural Networks},'' in
  \emph{IGARSS 2020-2020 IEEE International Geoscience and Remote Sensing
  Symposium}.\hskip 1em plus 0.5em minus 0.4em\relax IEEE, 2020, pp. 521--524.

\bibitem{ludeformsketchnet}
N.~G.~M. Lu, ``{DeformSketchNet: Deformable Convolutional Networks for Sketch
  Classification}.''

\bibitem{lai20213d}
S.~C. Lai, H.~K. Tan, and P.~Y. Lau, ``{3D Deformable Convolution for Action
  Classification in Videos},'' in \emph{International Workshop on Advanced
  Imaging Technology (IWAIT) 2021}, vol. 11766.\hskip 1em plus 0.5em minus
  0.4em\relax International Society for Optics and Photonics, 2021, p. 117660R.

\bibitem{sun2018integral}
X.~Sun, B.~Xiao, F.~Wei, S.~Liang, and Y.~Wei, ``{Integral Human Pose
  Regression},'' in \emph{Proceedings of the European Conference on Computer
  Vision (ECCV)}, 2018, pp. 529--545.

\bibitem{weng2018deformable}
J.~Weng, M.~Liu, X.~Jiang, and J.~Yuan, ``Deformable pose traversal convolution
  for 3d action and gesture recognition,'' in \emph{Proceedings of the European
  Conference on Computer Vision (ECCV)}, 2018, pp. 136--152.

\bibitem{bertasius2018object}
G.~Bertasius, L.~Torresani, and J.~Shi, ``{Object Detection in Video with
  Spatiotemporal Sampling Networks},'' in \emph{Proceedings of the European
  Conference on Computer Vision (ECCV)}, 2018, pp. 331--346.

\bibitem{mac2018locally}
K.-N.~C. Mac, D.~Joshi, R.~A. Yeh, J.~Xiong, R.~R. Feris, and M.~N. Do,
  ``{Locally-consistent deformable convolution networks for fine-grained action
  detection},'' \emph{arXiv preprint arXiv:1811.08815}, 2018.

\bibitem{9206944}
Z.~Li, H.~Pan, Y.~Zhu, and A.~K. Qin, ``{PGD-UNet: A Position-Guided Deformable
  Network for Simultaneous Segmentation of Organs and Tumors},'' in \emph{2020
  International Joint Conference on Neural Networks (IJCNN)}, 2020, pp. 1--8.

\bibitem{8999148}
M.~Pominova, E.~Kondrateva, M.~Sharaev, A.~Bernstein, S.~Pavlov, and
  E.~Burnaev, ``{3D Deformable Convolutions for MRI Classification},'' in
  \emph{2019 18th IEEE International Conference On Machine Learning And
  Applications (ICMLA)}, 2019, pp. 1710--1716.

\bibitem{Algorithm}
G.~Y. Huang~Q, Wang~D \emph{et~al.}, ``{Algorithm-hardware Co-design for
  Deformable Convolution},'' 2020, p. arXiv:2002.08357.

\bibitem{efficient}
S.~Ahn, J.-W. Chang, and S.-J. Kang, ``{An Efficient Accelerator Design
  Methodology For Deformable Convolutional Networks},'' in \emph{2020 IEEE
  International Conference on Image Processing (ICIP)}, 2020, pp. 3075--3079.

\bibitem{chen2016eyeriss}
Y.-H. Chen, J.~Emer, and V.~Sze, ``Eyeriss: A spatial architecture for
  energy-efficient dataflow for convolutional neural networks,'' in \emph{2016
  ACM/IEEE 43rd Annual International Symposium on Computer Architecture
  (ISCA)}.\hskip 1em plus 0.5em minus 0.4em\relax IEEE, 2016, pp. 367--379.

\bibitem{albericio2016cnvlutin}
J.~Albericio, P.~Judd, T.~Hetherington, T.~Aamodt, N.~E. Jerger, and
  A.~Moshovos, ``{Cnvlutin: Ineffectual-Neuron-Free Deep Neural Network
  Computing},'' \emph{ACM SIGARCH Computer Architecture News}, vol.~44, no.~3,
  pp. 1--13, 2016.

\bibitem{bromberger2016fpga}
M.~Bromberger, P.~Bastian, J.-P. Bergeest, C.~Conrad, V.~Heuveline, K.~Rohr,
  and W.~Karl, ``{FPGA-accelerated Richardson-Lucy Deconvolution for 3D Image
  Data},'' in \emph{2016 IEEE 13th International Symposium on Biomedical
  Imaging (ISBI)}.\hskip 1em plus 0.5em minus 0.4em\relax IEEE, 2016, pp.
  132--135.

\bibitem{wang2016deepburning}
Y.~Wang, J.~Xu, Y.~Han, H.~Li, and X.~Li, ``{DeepBurning: Automatic Generation
  of FPGA-based Learning Accelerators for the Neural Network Family},'' in
  \emph{Proceedings of the 53rd Annual Design Automation Conference}.\hskip 1em
  plus 0.5em minus 0.4em\relax ACM, 2016, p. 110.

\bibitem{zhang2018caffeine}
C.~Zhang, G.~Sun, Z.~Fang, P.~Zhou, P.~Pan, and J.~Cong, ``{Caffeine: Towards
  Uniformed Representation and Acceleration for Deep Convolutional Neural
  Networks},'' \emph{IEEE Transactions on Computer-Aided Design of Integrated
  Circuits and Systems}, 2018.

\bibitem{xu2021hyca}
D.~Xu, Q.~Wang, C.~Liu, C.~Chu, Y.~Wang, H.~Li, X.~Li, and K.-T. Cheng, ``Hyca:
  A hybrid computing architecture for fault tolerant deep learning,''
  \emph{arXiv preprint arXiv:2106.04772}, 2021.

\bibitem{xu2020hybrid}
D.~Xu, C.~Chu, Q.~Wang, C.~Liu, Y.~Wang, L.~Zhang, H.~Liang, and K.-T. Cheng,
  ``A hybrid computing architecture for fault-tolerant deep learning
  accelerators,'' in \emph{2020 IEEE 38th International Conference on Computer
  Design (ICCD)}.\hskip 1em plus 0.5em minus 0.4em\relax IEEE, 2020, pp.
  478--485.

\bibitem{xu2018fcn}
D.~Xu, K.~Tu, Y.~Wang, C.~Liu, B.~He, and H.~Li, ``{FCN-Engine: Accelerating
  Deconvolutional Layers in Classic CNN Processors},'' in \emph{Proceedings of
  the International Conference on Computer-Aided Design}.\hskip 1em plus 0.5em
  minus 0.4em\relax ACM, 2018, p.~22.

\bibitem{yazdanbakhsh2018ganax}
A.~Yazdanbakhsh, K.~Samadi, N.~S. Kim, and H.~Esmaeilzadeh, ``{Ganax: A Unified
  MIMD-SIMD Acceleration for Generative Adversarial Networks},'' in
  \emph{Proceedings of the 45th Annual International Symposium on Computer
  Architecture}.\hskip 1em plus 0.5em minus 0.4em\relax IEEE Press, 2018, pp.
  650--661.

\bibitem{yan2018gna}
J.~Yan, S.~Yin, F.~Tu, L.~Liu, and S.~Wei, ``Gna: Reconfigurable and efficient
  architecture for generative network acceleration,'' \emph{IEEE Transactions
  on Computer-Aided Design of Integrated Circuits and Systems}, vol.~37,
  no.~11, pp. 2519--2529, 2018.

\bibitem{zhang2017design}
X.~Zhang, S.~Das, O.~Neopane, and K.~Kreutz-Delgado, ``A design methodology for
  efficient implementation of deconvolutional neural networks on an fpga,''
  \emph{arXiv preprint arXiv:1705.02583}, 2017.

\bibitem{hegde2018morph}
K.~Hegde, R.~Agrawal, Y.~Yao, and C.~W. Fletcher, ``{Morph: Flexible
  Acceleration for 3D CNN-based Video Understanding},'' in \emph{2018 51st
  Annual IEEE/ACM International Symposium on Microarchitecture (MICRO)}.\hskip
  1em plus 0.5em minus 0.4em\relax IEEE, 2018, pp. 933--946.

\bibitem{wang2017enhanced}
H.~Wang, M.~Shao, Y.~Liu, and W.~Zhao, ``Enhanced efficiency 3d convolution
  based on optimal fpga accelerator,'' \emph{IEEE Access}, vol.~5, pp.
  6909--6916, 2017.

\bibitem{fan2018reconfigurable}
H.~Fan, H.-C. Ng, S.~Liu, Z.~Que, X.~Niu, and W.~Luk, ``Reconfigurable
  acceleration of 3d-cnns for human action recognition with block
  floating-point representation,'' in \emph{2018 28th International Conference
  on Field Programmable Logic and Applications (FPL)}.\hskip 1em plus 0.5em
  minus 0.4em\relax IEEE, 2018, pp. 287--2877.

\bibitem{RECOIN}
C.~Chu, F.~Chen, D.~Xu, and Y.~Wang, ``{RECOIN: A Low-Power Processing-in-ReRAM
  Architecture for Deformable Convolution},'' in \emph{GLSVLSI '21: Great Lakes
  Symposium on VLSI 2021}, 2021.

\bibitem{alwani2016fused}
M.~Alwani, H.~Chen, M.~Ferdman, and P.~Milder, ``{Fused-Layer CNN
  Accelerators},'' in \emph{2016 49th Annual IEEE/ACM International Symposium
  on Microarchitecture (MICRO)}.\hskip 1em plus 0.5em minus 0.4em\relax IEEE,
  2016, pp. 1--12.

\bibitem{samajdar2018scale}
A.~Samajdar, Y.~Zhu, P.~Whatmough, M.~Mattina, and T.~Krishna, ``{SCALE-Sim:
  Systolic CNN Accelerator Simulator},'' \emph{arXiv preprint
  arXiv:1811.02883}, 2018.

\bibitem{DRAM}
\url{https://www.micron.com/support/tools-and-utilities/power-calc}.

\bibitem{simonyan2014very}
K.~Simonyan and A.~Zisserman, ``{Very Deep Convolutional Networks for
  Large-Scale Image Recognition},'' \emph{arXiv preprint arXiv:1409.1556},
  2014.

\bibitem{badrinarayanan2017segnet}
V.~Badrinarayanan, A.~Kendall, and R.~Cipolla, ``{SegNet: A Deep Convolutional
  Encoder-Decoder Architecture for Image Segmentation},'' \emph{IEEE
  transactions on pattern analysis and machine intelligence}, vol.~39, no.~12,
  pp. 2481--2495, 2017.

\end{thebibliography}

\end{document}